\RequirePackage{fix-cm}
\documentclass[twocolumn,epjc3]{svjour3}  
\usepackage{booktabs} 

\smartqed  
\RequirePackage{graphicx}
\usepackage{url,hyperref} 
\usepackage[sorting=nty]{biblatex}
\usepackage{enumitem}
\usepackage{float}
\usepackage[most]{tcolorbox}
\usepackage{tikz}
\usepackage{pgfplots}
\usepackage{subcaption} 
\pgfplotsset{compat=1.18}

\journalname{Insert journal name here}

\usepackage{listings}
\usepackage{xcolor}

\usepackage{longtable}
\usepackage{booktabs}
\usepackage{cancel}
\usepackage{makecell}

\usepackage{textcomp}
\usepackage{booktabs}
\usepackage{lipsum}
\usepackage{multicol}
\usepackage{siunitx}

\usepackage{orcidlink}

\begin{document}

\title{
Empirical Evaluation of Large Language Models for Migration of Code Fragments to Post-Quantum Cryptography
}


\author{Javier Pallar\'{e}s~de~Bonrostro\thanksref{e1,addr1} \orcidlink{0009-0000-1442-7476}
        \and
        Ana~I. Gonz\'{a}lez-Tablas\thanksref{e2,addr1} \orcidlink{0000-0002-6259-8955}
        \and
        Mar\'{i}a~Isabel Gonz\'{a}lez~Vasco\thanksref{e3,addr2} \orcidlink{0000-0002-7452-9121} 
}

\thankstext{e1}{e-mail: javier.pallaresbonrostro@alumnos.uc3m.es}
\thankstext{e2}{e-mail: aigonzal@inf.uc3m.es}
\thankstext{e3}{e-mail: mariaigo@math.uc3m.es}

\authorrunning{Pallar\'{e}s, J.; Gonz\'{a}lez-Tablas, A.I. \and Gonz\'{a}lez~Vasco, M.I.} 

\institute{Computer Science and Engineering Department, Universidad Carlos III de Madrid, Spain \label{addr1}
           \and
Mathematics Department, Universidad Carlos III de Madrid, Spain \label{addr2}
}

\date{Received: date / Accepted: date}

\maketitle

\begin{abstract}
The transition to post-quantum cryptography (PQC) requires not only replacing vulnerable cryptographic primitives, but also refactoring the surrounding software logic that manages keys, parameters, serialization formats, and protocol-specific execution flows. While existing PQC migration frameworks provide organizational guidance, practical code-level remediation remains largely manual and error-prone. This paper evaluates whether large language models (LLMs) can be trained to assist in the migration of pre-quantum cryptographic code fragments to post-quantum or quantum-resistant counterparts while preserving functional correctness.

To this end, we introduce a reproducible experimental framework built around a synthetic yet executable dataset of 800 paired Python code fragments covering six cryptographic families---symmetric encryption, hashing, message authentication codes, authenticated encryption, digital signatures, and key exchange---as well as combined multi-primitive cases. Each pair is validated through category-specific functional tests, enabling both dataset quality control and objective evaluation of model-generated migrations. Four models are assessed: GPT-4.1 in a zero-shot setting, and fine-tuned versions of GPT-3.5-turbo, GPT-4.1-mini, and CodeLlama-7B-Instruct.

The results show that domain-specific fine-tuning is essential for reliable cryptographic migration. The fine-tuned GPT-4.1-mini model achieves the best overall performance, with a mean static similarity of 0.9072 and a dynamic functional correctness rate of 92.5\%, substantially outperforming the zero-shot baseline. A complementary validation on six open-source repositories further shows that the approach can produce useful migrations in localized cryptographic modules, while also revealing limitations in larger projects with complex dependencies and cross-module interactions. These findings suggest that fine-tuned LLMs can serve as practical components in future crypto-agile migration pipelines, provided they are coupled with automated verification and dependency-aware validation.

\end{abstract}

\section{Introduction}
\label{intro}

The upcoming global transition toward {Post-Quantum Cryptography (PQC)} represents one of the most complex challenges in modern cybersecurity and software engineering. Future quantum computers are expected to compromise the security of widely deployed asymmetric schemes such as RSA, Diffie–Hellman, ECDSA, and EdDSA through Shor’s algorithm, while Grover’s algorithm reduces the effective strength of symmetric primitives, forcing key sizes to increase in order to maintain equivalent security margins \cite{hasan2024framework}. Consequently, cryptographic infrastructures must migrate to quantum-resistant primitives such as the Module-Lattice-Based Key-Encapsulation Mechanism (ML-KEM) and the Module-Lattice-Based Digital Signature Algorithm (ML-DSA) digital-signature scheme, and adopt stronger symmetric constructions like AES-256 and SHA-3-512, in line with the migration roadmaps proposed by NIST, ENISA and the European Commission
for long-term protection in a post-quantum setting \cite{hasan2024framework,ENISA2021PQC, EU}.

{Beyond the challenges associated with adopting cryptographic constructions based on entirely new mathematical assumptions—including assessing the practical hardness of post-quantum problems and securely implementing these new primitives—the primary bottleneck in PQC adoption is operational.}
 Migration at scale requires discovering and refactoring cryptographic usage across large heterogeneous codebases, replacing obsolete primitives without breaking interoperability, and verifying that the resulting system preserves both its functional and security guarantees. Recent migration frameworks and guidance documents emphasise that, in the absence of {crypto-agility}, these processes tend to be slow, costly and error-prone, and therefore call for systematic inventory, dependency analysis and staged replacement strategies \cite{hasan2024framework,sikeridis2023elca,ENISA2021PQC}. The motivation for this study arises from this gap: although such conceptual frameworks exist, there is a lack of practical tools capable of performing large-scale cryptographic refactoring automatically. This work addresses that gap from a software-engineering perspective, leveraging synthetic data, LLM-based migration models, and functional validation to automate the transition toward PQC.

At the same time, {Large Language Models (LLMs)} have reached remarkable maturity in code understanding, automated refactoring and repository-scale modernization. Industrial and academic studies demonstrate that LLMs can perform coordinated transformations across thousands of files, generate unit tests and validate behavioral equivalence, often reducing manual effort by up to 90 \% \cite{ziftci2025migrating,shirafuji2023refactoring,mahmud2024automated,zhou2023hybrid}. Surveys on LLMs for vulnerability detection and secure coding further highlight their increasing ability to reason about program semantics and security contexts \cite{ijis2025survey}. This convergence between an urgent need (massive PQC migration) and an emerging capability (AI-assisted code transformation) opens an opportunity to explore whether LLMs can accelerate and help systematize post-quantum transitions in a reliable and verifiable way.

For cryptographic software, recent evidence suggests that LLMs are beginning to outperform or complement static tools in specialized security tasks. Masood et al. \cite{masood2024llmcryptoguard} compare multiple LLMs with CryptoGuard, CogniCrypt and Snyk Code on OWASP and MASC benchmarks, showing that GPT-4-o-mini achieves higher precision and recall in detecting API misuses and weak configurations. Li et al. introduce CryptoScope \cite{cryptoscope2025}, which combines retrieval-augmented generation (RAG) and chain-of-thought prompting over a curated 12,000-entry cryptographic knowledge base to identify logic-level vulnerabilities. Maskey et al. \cite{maskey2025llmcrypto} evaluate LLMs on cryptanalysis and side-channel reasoning tasks, highlighting both capability boundaries and dual-use risks. These studies demonstrate that LLMs can reason about cryptographic constructs; yet, no work has established a complete pipeline for verified code migration toward PQC.

Meanwhile, ecosystem-scale measurements reveal that most deployed applications remain non-ready for PQC. Strauss et al. \cite{strauss2025quantumreadiness} analyze more than 4,000 Android apps, uncovering widespread use of SHA-1, MD5, and RSA, and assess whether state-of-the-art LLMs can assist in replacing them. Their experiments show partial success in local substitutions (for example, SHA-1 → SHA-256) but systematic failure in end-to-end PQC migrations such as RSA/ECC → Kyber/Dilithium due to missing multi-file context, dependency resolution, and validation. Far from being a negative result, this study provides a realistic baseline that clarifies where progress is required: semantic coherence across files, explicit cryptographic context, and automated functional verification.

Taken together, these developments motivate the central question of this paper:
\begin{quote}
\emph{Can large language models be effectively trained and evaluated to assist in the migration of pre-quantum cryptographic code to post-quantum counterparts while preserving functional correctness?}
\end{quote}

To answer this question, we design an experimental framework based on synthetic data generation, automatic validation and LLM fine-tuning, enabling a controlled and reproducible evaluation of AI-assisted cryptographic migration.

\subsection{Related Work}

\paragraph{Migration frameworks and crypto-agility.}
Hasan et al.\ \cite{hasan2024framework} propose a structured framework for migrating legacy systems to PQC through dependency analysis and case studies, while Sikeridis et al.\ \cite{sikeridis2023elca} extend the concept of crypto-agility to enterprise governance, introducing ELCA as a model for rapid algorithm substitution. These works define the organisational process for PQC transition but stop short of automating code transformation. Complementary initiatives such as IBM’s Cryptography Bill of Materials (CBOM) and CISA’s Automated Cryptographic Discovery and Inventory (ACDI) focus primarily on discovery and inventory rather than remediation and code refactoring \cite{ibm2024cbom,cisa2024pqcstrategy}. 
{In parallel, official roadmaps by ENISA and NIST explicitly recommend proactive migration away from RSA/ECC towards standardised post-quantum schemes, including the adoption of hybrid classical–post-quantum (C-PQC) constructions, while emphasising the need to incorporate crypto-agility into existing architectures,} rather than treating PQC as a one-off switch \cite{ENISA2021PQC,nist-sp800-232-ipd, nist_cswp_39_2025}.

\paragraph{LLMs for code refactoring and modernization.}
LLM-driven refactoring research has grown rapidly. Google reports multi-file code edits performed autonomously by LLM agents within its monorepository \cite{ziftci2025migrating}. Tools such as GUPPY and Hybrid API Migration demonstrate automated API upgrades on Android using prompt-based generation and semantic patch refinement \cite{mahmud2024automated,zhou2023hybrid}. Models such as TransCoder and Code-LLaMA achieve high translation accuracy between programming languages \cite{lachaux2020unsupervised,roziere2023code}, while Few-Shot Program Refactoring attains almost 96 \% functional correctness in structural improvements with GPT-3.5 \cite{shirafuji2023refactoring}. Despite these achievements, all operate in security-agnostic domains and do not consider cryptographic correctness or protocol integrity.

\paragraph{LLMs in security and cryptographic reasoning.}
Recent surveys \cite{ijis2025survey} and benchmarks \cite{masood2024llmcryptoguard,cryptoscope2025,maskey2025llmcrypto} confirm that LLMs can detect cryptographic misuse and reason about algorithmic intent, bridging part of the gap between static analysis and human auditing. However, the ability to produce and verify secure cryptographic replacements remains untested. These works therefore provide context and motivation for exploring LLMs not only as detectors but as active participants in secure code transformation.

\paragraph{Ecosystem readiness and empirical studies.}
Strauss et al. \cite{strauss2025quantumreadiness} represent the first attempt to use LLMs for real-world PQC migration. Their findings show that the models succeed only in trivial substitutions and fail when context and validation are required. Other empirical efforts address complementary angles: Ahmed et al. \cite{ahmed2025pqc} survey PQC support in cryptographic libraries; Ricchizzi et al. \cite{ricchizzi2025industrial} examine industrial deployment of hybrid certificates; and policy roadmaps from NIST and the PQC Coalition \cite{nist2025whitepaper,pqcc2025roadmap} outline strategic but non-automated migration guidance. None provide reproducible, code-level experiments assessing LLM capability.

Overall, prior literature provides three partial pillars: (1) frameworks and policies for PQC migration planning, (2) advances in LLM-based software refactoring, and (3) empirical evidence of ecosystem unpreparedness. What is missing is a controlled environment that connects these strands and quantifies how well LLMs can perform verified cryptographic migration tasks.

\subsection{Contributions}

This work proposes a reproducible experimental framework that bridges conceptual PQC migration models and practical, verifiable LLM-assisted transformation at the source-code level. The contributions are as follows:

\begin{enumerate}[label=\textbf{C\arabic*.}]
    \item \textbf{Dataset for reproducible cryptographic migration.}  
    We introduce a balanced and verifiable corpus of approximately 800 pre- to post-quantum code pairs across seven categories: hashing, MAC, digital signatures, key-exchange / KEM, authenticated encryption, and combined schemes. Each pair includes unit tests (e.g., {sign/verify}, {encaps/decaps}, {encrypt/decrypt}) to ensure correctness. To our knowledge, this is the first structured dataset designed specifically for evaluating LLMs on PQC migration tasks. The dataset has been published as an openly available research artifact to support reproducibility and future benchmarking \cite{7GK4MJ_2025}. To our knowledge, this is the first structured dataset designed specifically for evaluating LLMs on PQC migration tasks.

    \item \textbf{LLMs trained for verifiable cryptographic migrations.}  
    Four representative models (GPT-4.1, GPT-3.5-turbo, GPT-4.1-mini and CodeLlama-7B) are evaluated under zero-shot and fine-tuned configurations. Metrics combine lexical-structural similarity, functional success rate, execution time and cost per token. Fine-tuned models show significant improvement in localized, verifiable migrations such as HMAC-SHA1 to HMAC-SHA3-256 or AES-128 to AES-256, while complex PQC replacements like RSA to Kyber remain challenging.

    \item \textbf{Automated verification pipeline.}  
    We implement a validation pipeline integrating static analysis, dynamic testing and runtime verification to confirm that migrated code is syntactically valid, executes and produces valid cryptographic outputs. This approach transforms PQC migration into a measurable benchmark, enabling objective comparison across models and full reproducibility.

    \item \textbf{Alignment with enterprise frameworks and crypto-agility.}  
    The workflow maps its stages (inventory, candidate selection, patch generation and validation) to existing PQC migration frameworks \cite{hasan2024framework,sikeridis2023elca}, facilitating its integration into organizational change-management and continuous-validation processes.
\end{enumerate}

Together, these contributions complement large-scale empirical work such as Strauss et al. \cite{strauss2025quantumreadiness}. While that study documents the current limits of LLMs in practical PQC migration, our framework provides the experimental foundation --data, methodology and verification-- to train and evaluate such models under reproducible conditions. The combination of empirical observation and controlled experimentation establishes a basis for future research on reliable, AI-assisted post-quantum migration.

\subsection{Structure of the paper}

The remainder of this paper is organized as follows. Section~\ref{sec:background-llm-pipelines} provides the necessary background on large language models and software migration pipelines, with a particular focus on their relevance to cryptographic transformation tasks. Section~\ref{sec:methodology} presents the overall methodology of the study. Section~\ref{sec:dataset-generation} describes the dataset construction process, including primitive selection, synthetic generation, and validation. Next, Section~\ref{sec:LLM_general} details the experimental setup for model selection, preparation, fine-tuning, and testing. Section~\ref{sec:evaluation} introduces the automated evaluation framework used to assess both static similarity and dynamic functional correctness. The quantitative results are reported in Section~\ref{sec:results}, followed by the application of the best-performing model to real-world repositories in Section~\ref{sec:real-world}. Finally, Section~\ref{sec:discussion-limitations} discusses the main limitations of the approach, and Section~\ref{sec:conclusions} concludes the paper and outlines future research directions. Additional details on dataset composition are provided in ~\ref{app:dataset}.

\section{Background on LLMs and migration pipelines}
\label{sec:background-llm-pipelines}

Large language models (LLMs), built on the Transformer architecture introduced by Vaswani et al. \cite{vaswani2017attention}, have rapidly evolved into general-purpose systems for code understanding, generation, and transformation. Recent surveys confirm their growing impact across software engineering tasks, including code generation, refactoring, testing, repair, and repository-scale maintenance \cite{zhang2024surveylargelanguagemodels,jiang2024surveycodellm}. Industrial studies further show that these models can apply thousands of coordinated edits across monolithic repositories with minimal human supervision. For instance, Ziftci et al.\ report that LLM-driven agents were able to execute more than 93,000 coherent modifications within Google's monorepository, maintaining semantic consistency and passing the full battery of automated tests \cite{ziftci2025migrating}. Similar results appear in refactoring and API modernisation tasks, where LLMs generate patches, run tests, and refine outputs iteratively \cite{mahmud2024automated,zhou2023hybrid}. Academic work on program translation further demonstrates that models such as TransCoder \cite{lachaux2020unsupervised} and Code LLaMA \cite{roziere2023code} can perform unsupervised or instruction-tuned translation between programming languages with high structural fidelity. Few-shot refactoring experiments have also shown that models like GPT-3.5 can achieve more than 95\% correctness when guided with a small set of examples \cite{shirafuji2023refactoring}.

These advances have led to the adoption of {pipeline oriented} workflows in software engineering. In such pipelines, the LLM is only one component inside a broader chain that includes discovery, patch generation and validation. A common pattern is: (i) static or semantic analysis identifies candidate locations; (ii) the LLM synthesises the patch; and (iii) compilation, test execution or fuzzing validates the result. This structure mirrors modern CI/CD processes and generalises across domains such as Android API upgrades \cite{mahmud2024automated,zhou2023hybrid}, multi file transformations in large codebases \cite{ziftci2025migrating}, and even multimodal repository reconstruction \cite{lin2025autop2c}. The same organisation is now standard in scientific applications that require reproducible train/test cycles and controlled comparisons across multiple model families.

One of the design dimensions of LLM pipelines is the exploitation mode. 
The most popular ones are described next:
\begin{itemize}
\item {Zero shot prompting} treats the model as a general assistant: the input code and a natural language instruction describe the target transformation. This requires no model adaptation but performs poorly in specialised domains such as cryptography, where subtle type or parameter constraints are essential. 
\item {Retrieval augmented generation (RAG)} improves grounding by injecting external documentation or specifications retrieved from a vector index; this reduces hallucinations and enables adherence to strict policies, an approach used in hybrid pipelines such as CryptoScope \cite{cryptoscope2025}. 
\item Finally, {fine tuning} adapts the model weights to a domain specific dataset, achieving the highest task accuracy. This strategy underpins advanced refactoring frameworks and translation systems \cite{lachaux2020unsupervised,roziere2023code,shirafuji2023refactoring} and is the most suitable option when correctness and reproducibility are required.
\end{itemize}

A second design dimension concerns the number of models involved. Recent work shows that multi stage or agent based systems outperform single model approaches. Google’s migration agent coordinates planning, patching and validation steps \cite{ziftci2025migrating}, while AutoP2C decomposes tasks into extraction, hierarchical planning and iterative repair modules \cite{lin2025autop2c}. This modular design allows systematic evaluation: each model (classifier, generator, reviewer) can be swapped or ablated while keeping the pipeline intact.

\paragraph{Cryptographic code migration pipelines. }
Despite this progress, cryptographic migration remains under explored. Existing research focuses largely on detecting misuses (such as API misuse detection in LLM based analyses \cite{masood2024llmcryptoguard} or logic level vulnerability identification \cite{cryptoscope2025}) rather than generating verified replacements for complete primitives. There are no prior works addressing end to end migration from RSA or ECDSA to post quantum schemes such as Kyber or Dilithium, nor evaluating these transformations functionally. This gap motivates the present study, which adopts the pipeline perspective described above and integrates dataset construction, fine tuning and dynamic testing to examine how far current LLMs can be pushed as practical engines for post quantum cryptographic migration.

\section{Methodology}
\label{sec:methodology}

The motivation for this study stems from the current lack of practical tools capable of performing large-scale cryptographic migrations. Although several frameworks have been proposed to guide post-quantum readiness at an organizational level \cite{hasan2024framework,sikeridis2023elca,nist2025whitepaper,pqcc2025roadmap}, these approaches remain conceptual and depend heavily on manual auditing to identify, modify and validate the affected cryptographic components. Existing inventory and discovery initiatives, such as IBM’s CBOM and CISA’s ACDI \cite{ibm2024cbom,cisa2024pqcstrategy}, assist in locating legacy primitives but do not automate their remediation. Furthermore, recent empirical studies show that current LLMs can detect cryptographic misuses but do not reliably achieve end-to-end post-quantum migrations in real software systems \cite{masood2024llmcryptoguard,cryptoscope2025,strauss2025quantumreadiness}.

\begin{figure}
    \centering
    \includegraphics[width=1\linewidth]{ 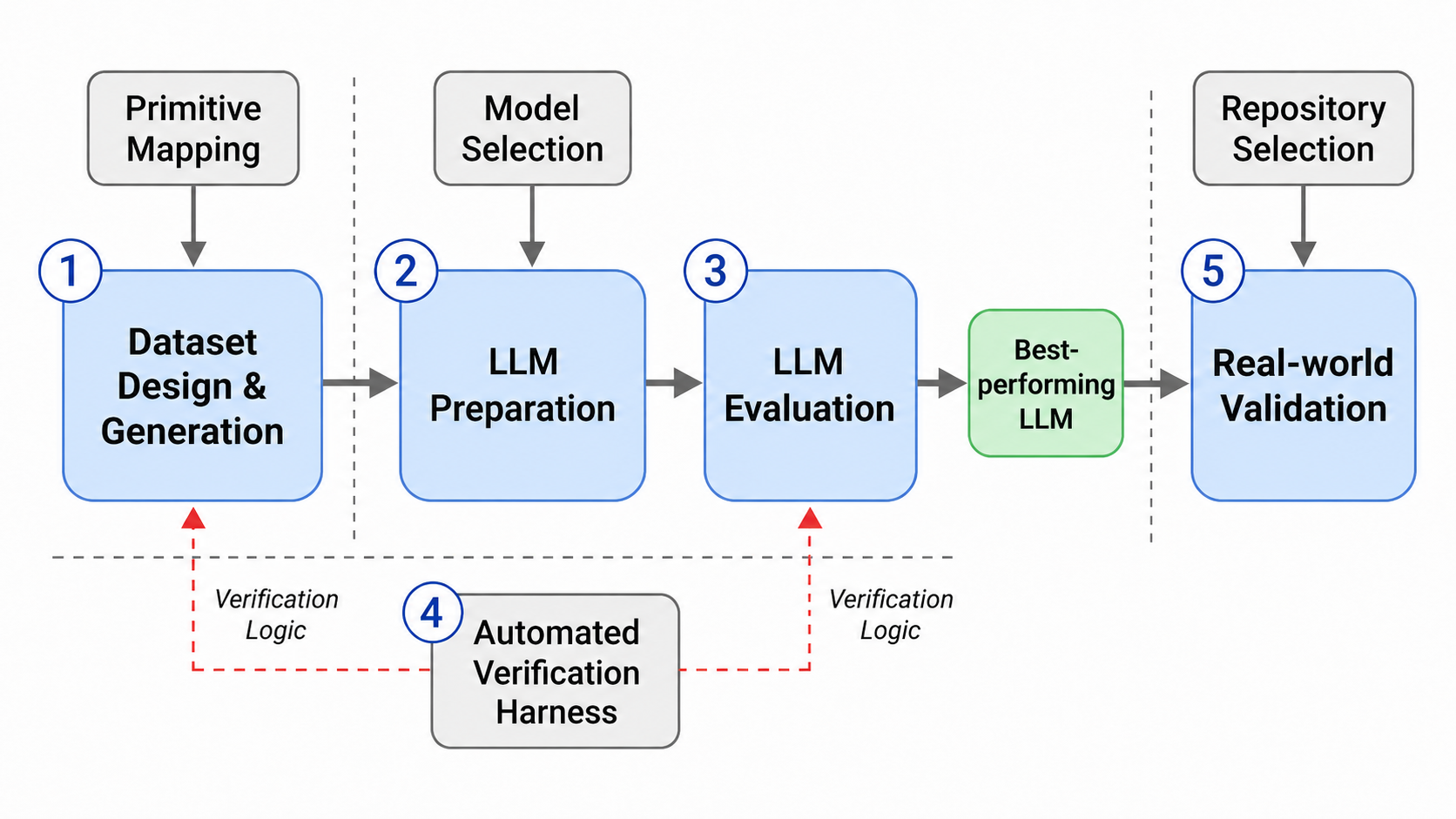}
    \caption{Overall methodological pipeline of the study. The workflow is organized into four main stages: dataset design and generation, LLM preparation (training), LLM evaluation (testing), and real-world validation. The figure also highlights the dual role of the automated verification harness, which is reused both to validate the dataset during construction and to evaluate model-generated migrations. } 
    \label{fig:methodology}
\end{figure}

The objective of this research is to evaluate whether fine-tuned large language models (LLMs) can automatically and reliably transform pre-quantum cryptographic code into functionally equivalent post-quantum implementations, minimizing manual intervention and human error. 
We have designed the methodology as a reproducible, data-driven experimental process, addressing the problem from a software-engineering perspective and following a modular experimental design that includes synthetic data generation and LLM training and validation (see Figure \ref{fig:methodology}). 

\paragraph{Dataset design and generation.} A central component of the methodology is the construction of a dedicated dataset for post-quantum code migration (see phase \textcircled{\small{1}} in Fig.~\ref{fig:methodology}). To the best of our knowledge, no publicly available benchmark currently provides paired pre-quantum and post-quantum implementations designed specifically for training and evaluating LLMs on this task. Moreover, collecting such pairs directly from real-world repositories is difficult in practice, since migrations are often incomplete, distributed across multiple files, tightly coupled to project-specific dependencies, or simply unavailable in aligned pre/post form. For this reason, we opted for a synthetic yet functionally accurate dataset, which makes it possible to generate large, balanced, and controlled examples while avoiding the unpredictability of production repositories. This design supports systematic evaluation across cryptographic categories such as hashing, message authentication, digital signatures, symmetric encryption, authenticated encryption, and combined cases, while preserving the verifiability needed for automated testing. As a result, the dataset serves as a central methodological asset for both model training and controlled evaluation.

\paragraph{LLM training and testing.} From a technical point of view, the LLM preparation and evaluation stages (see phases \textcircled{\small{2}}  and\textcircled{\small{3}} in Fig.~\ref{fig:methodology}) frame the migration of cryptographic primitives as a transformation problem between two domains: quantum vulnerable algorithms such as RSA, ECDSA  on one side, and quantum-safe primitives like CRYSTALS–Kyber or CRYSTALS–Dilithium on the other. However, the difficulty lies not in the mathematical substitution itself, but in the contextual dependencies that surround each primitive. Legacy codebases often integrate cryptography through indirect abstractions, custom wrappers or dynamic imports, which hinder automated detection and replacement.

\paragraph{Automated evaluation of code migration.}
The central methodological idea in this work is to treat code migration as a \emph{verifiable generation task}. Rather than relying exclusively on static analysis, the model receives a concrete code fragment as input and must output a modified version that preserves functionality while replacing quantum-vulnerable cryptographic elements with their post-quantum counterparts. 
Crucially, the difficulty lies less in the mathematical substitution itself than in the surrounding software context: key material handling, parameter initialization, encoding and serialization conventions, error and exception management, and the interaction with application-specific wrappers or helper APIs. 
Accordingly, the goal is not only to generate syntactically valid code, but to produce changes that remain executable and behavior-preserving under realistic usage patterns.

To operationalize this notion of verifiability, the automated verification harness (see phase \textcircled{\small{4}}  in Fig.~\ref{fig:methodology}) is applied at two points of the pipeline: during dataset design and generation (phase \textcircled{\small{1}} ), to ensure that the constructed examples are functionally valid, and during LLM evaluation (phase \textcircled{\small{3}}), to assess whether the migrated outputs preserve the expected behavior.
These tests check that the migrated code can be executed and that the expected input--output behavior is preserved (e.g., decrypting yields the original plaintext, signature verification succeeds for valid messages and fails for tampered ones), providing an objective signal of migration correctness beyond surface-level similarity.

\paragraph{Application to real-world cases.} Finally, to close the methodological loop, the fine-tuned model was applied to real open-source projects containing genuine cryptographic dependencies (see phase\textcircled{\small{5}} in Fig.~\ref{fig:methodology}). This final phase allowed assessing the system’s practical usefulness beyond controlled settings, providing insight into how LLM-assisted migration behaves under realistic complexity, modularity and dependency constraints. 

\paragraph{Automated pipeline.}
Additionally, the workflow from dataset preparation to model evaluation was implemented as a scripted pipeline to support reproducibility and traceability. Automation primarily concerns the repeatable execution of (i) dataset formatting and split generation (e.g., conversion to ChatML and stratified train/validation splits), (ii) inference and static evaluation on validation samples, and (iii) functional evaluation through an automated test harness that executes the corresponding base-case validators and aggregates pass/fail summaries.
At the same time, key design steps remain human-guided, such as defining base cases, curating the set of variations included in the final dataset, configuring training runs through notebooks, and manually resolving failures when applying migrations beyond the controlled benchmark. 
Across stages, results are recorded as JSON/JSONL artifacts (e.g., per-sample similarity and per-test summaries), enabling quantitative comparisons across models.

\paragraph{Language and cryptographic library selection.}
All experiments were conducted in Python because it provides a practical balance between cryptographic expressiveness, tooling maturity, and experimental reproducibility. Other languages such as C/C++ and Java are also widely used in security-critical software and would be natural candidates for production-oriented PQC migration studies. However, Python offers several advantages for a controlled experimental setting: it has a mature cryptographic ecosystem, including \texttt{PyCA/cryptography}, direct access to post-quantum primitives through \texttt{liboqs-python}, and a simple execution model that facilitates dynamic testing of generated snippets. These properties make it easier to construct self-contained examples, execute them inside an automated test harness, and compare model outputs under repeatable conditions.

The choice of Python should therefore be understood as a methodological decision rather than a claim that PQC migration is only relevant to Python codebases. By reducing language-level complexity, the experiments can focus on the core research question: whether LLMs can learn to transform cryptographic logic while preserving functional correctness. Extending the same methodology to other widely deployed languages such as C/C++, Java, Go, or Rust is left as future work.

\section{Dataset design and generation}
\label{sec:dataset-generation}

\begin{figure}
    \centering
    \includegraphics[width=1\linewidth]{ 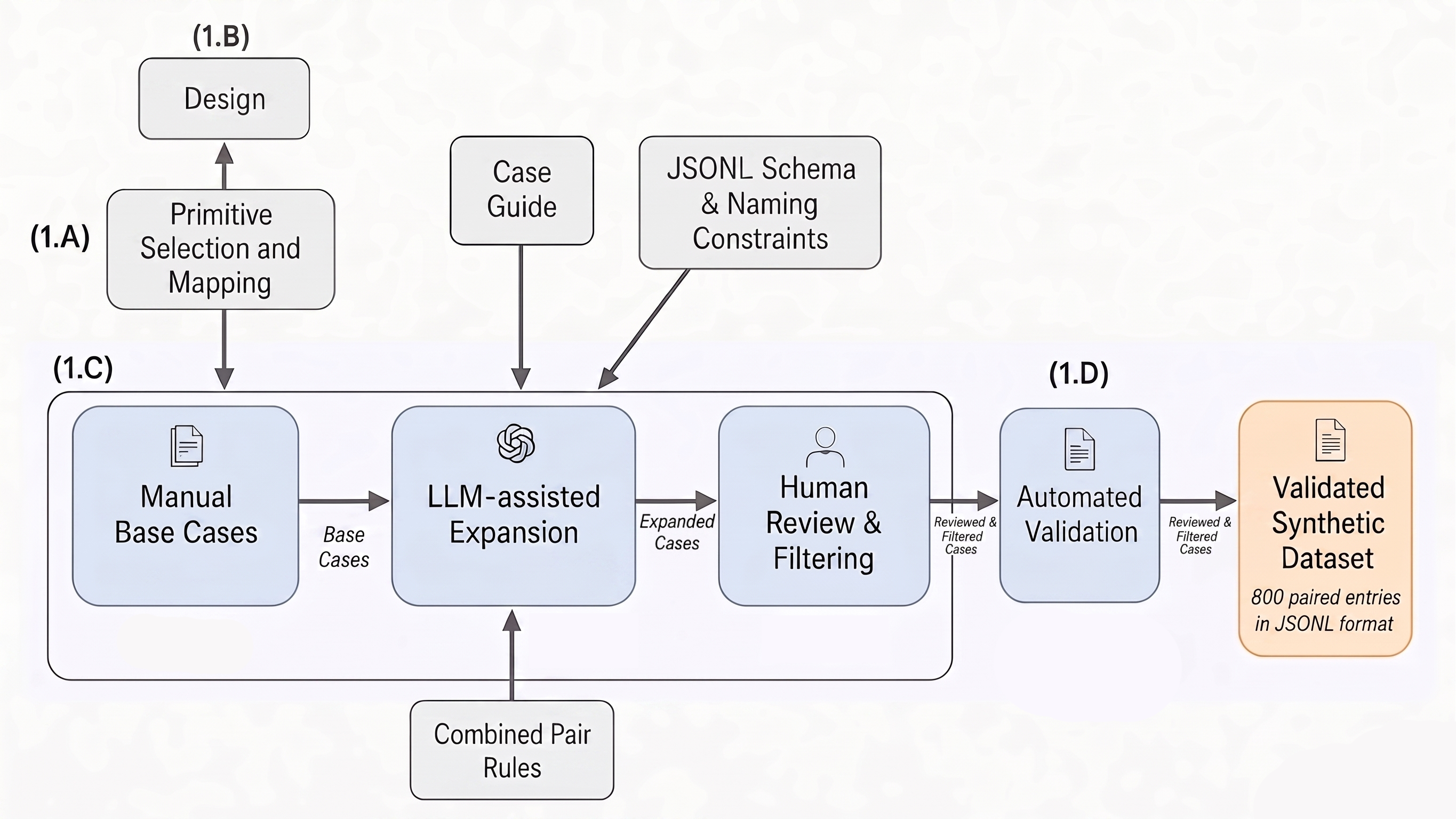}
    \caption{\textit{Dataset Generation Phase}. The internal flow starts with the manual creation of migrations, continues with synthetic generation and validation of the resulting corpus.}
    \label{fig:dataset_method}
\end{figure}

The overall construction process of the corpus is summarized in Fig.~\ref{fig:dataset_method}, while Tables~\ref{tab:primitive-migration-summary} and~\ref{tab:combined-pair-distribution} describe the primitive mappings and the combined cases represented in the final dataset.

\begin{table*}[t]
\centering
\renewcommand{\arraystretch}{1.3}
\caption{Summary of primitive migrations and estimated difficulty per category.}
\label{tab:primitive-migration-summary}

\begin{tabular}{|p{3cm}|p{3cm}|p{3cm}|p{5.2cm}|}
\hline
\textbf{Category} & \textbf{Quantum Vulnerable Crypto} & \textbf{Quantum Resistant Crypto} & \textbf{Migration difficulty estimated} \\
\hline
Symmetric encryption & 3DES, AES-128 & AES-256 &
\textbf{Medium}. Duplicated padding, mis-sized keys, obsolete imports. \\
\hline

Hash functions & MD5, SHA-1/2, BLAKE2b/s & SHA3-256 &
\textbf{Low–Medium}. Missing algorithms, reused hash objects, identifier mismatches. \\
\hline

Message authentication codes (MAC) & CMAC, HMAC-SHA1/2, BLAKE2, JWT-HMAC & HMAC-SHA3-256 &
\textbf{Medium}. Deprecated imports, method-signature errors, mismatched generation/verification. \\
\hline
Authenticated encryption & AES-GCM, AES-CCM, AES-OCB3, AES-SIV, Fernet & ChaCha20-Poly1305 &
\textbf{Medium}. Invalid key sizes, missing methods, incorrect class wiring. \\
\hline

Digital signatures & RSA, DSA, ECDSA, Ed25519, Ed448 & Dilithium5, SPHINCS+ &
\textbf{High}. Key-order errors, padding inconsistencies, incoherent signatures, wrong algorithm selection. \\
\hline

Key exchange & DH-2048/3072, ECDH, X25519, X448 & Kyber1024 &
\textbf{High}. Non-shared secrets, incorrect method signatures, asynchronous API mismanagement. \\
\hline

\end{tabular}

\end{table*}

\begin{table*}[t]
\centering
\renewcommand{\arraystretch}{1.3}
\caption{Distribution of combined pair cases in the dataset. Values indicate the number of variations for each unordered pair of primitive families.}
\label{tab:combined-pair-distribution}

\begin{tabular}{|p{3.4cm}|>{\centering\arraybackslash}p{2.2cm}|>{\centering\arraybackslash}p{2.1cm}|>{\centering\arraybackslash}p{1.7cm}|>{\centering\arraybackslash}p{3.0cm}|>{\centering\arraybackslash}p{2.8cm}|}
\hline
\textbf{Primitive family} &
\makecell{\textbf{Digital}\\\textbf{signatures}} &
\makecell{\textbf{Key}\\\textbf{exchange}} &
\makecell{\textbf{Hashing}} &
\makecell{\textbf{Message}\\\textbf{authentication}\\\textbf{codes}} &
\makecell{\textbf{Authenticated}\\\textbf{encryption}} \\
\hline

\textbf{Symmetric encryption} & 16 & 16 & 16 & 8 & 16 \\
\hline
\textbf{Digital signatures} & - & 16 & 16 & 8 & 16 \\
\hline
\textbf{Key exchange} & - & - & 16 & 8 & 16 \\
\hline
\textbf{Hashing} & - & - & - & 8 & 16 \\
\hline
\textbf{Message authentication codes} & - & - & - & - & 8 \\
\hline

\multicolumn{5}{|r|}{\textbf{Total number of variations}} & 200 \\
\hline
\end{tabular}

\end{table*}

\paragraph{Primitive selection and migration mapping.}
The first step in the dataset construction process is primitive selection and migration mapping (see 1.A in Fig.~\ref{fig:dataset_method}). The process requires a consistent mapping between pre-quantum and post-quantum primitives so that the resulting dataset reflects realistic software refactoring scenarios rather than purely theoretical substitutions. This mapping is guided by recommendations from NIST, ENISA, and recent academic analyses of post-quantum transition strategies \cite{hasan2024framework,ENISA2021PQC,sikeridis2023elca}. The objective is not only to replace cryptographic algorithms that are vulnerable in the quantum setting, but also to capture the structural changes that arise when moving from classical schemes such as RSA or ECDH to lattice-based constructions such as Dilithium or Kyber.

At a methodological level, the key assumption is that post-quantum cryptography differs from classical cryptography not only in the underlying mathematics, but also in interface design, key material structure, and operational flow. Some migrations involve relatively local changes, such as increasing symmetric key sizes, whereas others require replacing entire workflows, as in the transition from ECDSA to Dilithium or from ECDH to Kyber, where the involved primitives rely on fundamentally different algebraic structures and therefore demand substantially different arithmetic implementations.
These differences directly shape the complexity of the dataset and the kinds of software-engineering errors that the models must learn to avoid.

Symmetric cryptography provides the most straightforward migration path. Under Grover's model, the effective security of a symmetric cipher is reduced, but robustness can be preserved by increasing the key size \cite{hasan2024framework}. For this reason, AES-256 is selected as the standard post-quantum target for symmetric encryption. Hash functions follow a similar logic: while SHA-2 remains viable with sufficiently long outputs, SHA-3 offers a more modern structure and stronger security margin \cite{dworkin2015sha}. Accordingly, SHA3-256 is used as the default target for hashing tasks, while SHA3-512 is reserved for higher-assurance contexts.

Message authentication is considered in two settings: standalone MAC constructions and AEAD schemes, where integrity is integrated into the encryption primitive. Some MAC constructions, such as CMAC and GMAC, become problematic under quantum-superposition adversaries \cite{kaplan2016breaking,lang2023post}, whereas HMAC remains secure when instantiated with a quantum-resistant hash. For this reason, HMAC--SHA3-256 is selected as the canonical post-quantum MAC. Authenticated encryption requires separate treatment. Although AES-GCM is widely deployed, it inherits structural issues from GMAC in quantum settings \cite{lang2023post}. ChaCha20--Poly1305, by contrast, does not exhibit these weaknesses and is therefore used as the target AEAD scheme in the dataset.

Public-key primitives require deeper structural changes. All classical schemes based on factoring or discrete logarithms are broken by Shor's algorithm \cite{hasan2024framework}, and ENISA explicitly identifies them as high-priority migration targets \cite{ENISA2021PQC}. For digital signatures, CRYSTALS--Dilithium is selected as the primary post-quantum target due to its robustness and its role in the NIST standardization process. The dataset uses Dilithium5 to remain aligned with the highest security level considered in the study. In addition, selected signature examples also include SPHINCS+, a stateless hash-based signature scheme standardized by NIST, in order to cover an alternative post-quantum signature family with different security assumptions. For key establishment, CRYSTALS--Kyber is selected, specifically the Kyber1024 variant, which corresponds to the strongest parameter set in the Kyber family \cite{hasan2024framework,sikeridis2023elca, nist2022quantum}.

\textit{Standardization note.} NIST has standardized Kyber, Dilithium and SPHINCS+ as ML-KEM, ML-DSA and SLH-DSA, respectively\cite{nist2024fips203,nist2024fips204,nist2024fips205}.  However, the experimental pipeline used in this work relies on the CRYSTALS-family identifiers exposed by the pinned \texttt{liboqs}/\texttt{liboqs-python} toolchain. This choice does not alter the methodological conclusions of the study, since moving from the submission identifiers to the standardized ones is mainly an implementation-level substitution rather than a change in the migration problem itself. Repeating the experiments with ML-KEM, ML-DSA and SLH-DSA identifiers is left as future work.

Table~\ref{tab:primitive-migration-summary} summarizes the selected pre-quantum and post-quantum mappings for the main primitive families, together with the estimated migration difficulty.

\paragraph{Dataset design.}
A central contribution of this work is the construction of a controlled dataset specifically designed to train and evaluate large language models on cryptographic migration tasks (see 1.B in Fig.~\ref{fig:dataset_method}). The dataset acts both as a benchmark and as a source of supervised fine-tuning examples, enabling model behavior to be measured under repeatable and verifiable conditions. Its design follows one main principle: every entry must be functionally self-contained and executable, so that correctness can be assessed automatically.

The final dataset contains 800 entries. Each entry consists of a paired transformation from a pre-quantum implementation to its post-quantum counterpart. Of these, 600 entries correspond to single-family migrations distributed across six cryptographic categories (listed in Table~\ref{tab:primitive-migration-summary}): symmetric encryption, digital signatures, key exchange, hashing, message authentication codes, and authenticated encryption. 

The remaining 200 entries correspond to \emph{combined} cases.
In this work, \emph{combined} refers specifically to snippets in which exactly two different cryptographic primitive families appear together and must be migrated coherently within the same code fragment. This restriction is intentional: pairwise combinations introduce realistic multi-primitive interactions while keeping each migration concise, interpretable, and functionally testable. Larger compositions involving three or more primitives would increase realism, but would also add noise and make it harder to attribute failures to a specific migration pattern. Extending the dataset to more complex multi-primitive compositions is left as future work. 

The combined subset was constructed by considering all unordered pairwise combinations among the six primitive families in the dataset: symmetric encryption, authenticated encryption, hashing, MAC, digital signatures, and key exchange. This results in 15 possible pair categories. The final subset contains 200 combined variations: most pair categories contain 16 variations, while the five pair categories involving MAC contain 8 variations each.

This allocation reflects a pragmatic corpus-design decision. All pairwise interactions are represented, but MAC-involving combinations are sampled more compactly because they tend to share a stable tag-generation/tag-verification structure and therefore introduce less additional structural variability than combinations involving encryption, key establishment, or digital signatures. This keeps the combined subset bounded while still exposing the model to authentication logic in multi-primitive scenarios. The resulting distribution is summarized in Table~\ref{tab:combined-pair-distribution}.

A key design decision was to ensure that every example captures the complete functional flow associated with its cryptographic family. In practice, this means that symmetric encryption examples include both encryption and decryption, digital signature examples include signing and verification, key-establishment examples validate the full shared-secret establishment flow (encapsulation/decapsulation for KEM-based schemes and shared-secret derivation for classical DH/ECDH-style schemes), MAC examples include tag generation and verification, and hashing examples include digest computation together with the corresponding validation logic used in the test harness. This constraint ensures that migration is evaluated at the level of executable cryptographic behavior rather than as a superficial API substitution.

Each dataset entry is stored in JSON Lines (JSONL) format, which is commonly used for LLM training corpora \cite{de2024nos_corpusnos}. In addition to the paired pre- and post-quantum code fragments, each entry contains metadata such as the base case identifier, variation identifier, migration category, and the source and target algorithms involved. This structure supports downstream parsing, tokenization, and evaluation while preserving a uniform representation across all categories. The schema used throughout the dataset is shown in Listing~\ref{lst:jsonl_schema}, which formalizes the structure followed by all manually created and automatically expanded entries.

\begin{lstlisting}[basicstyle=\ttfamily\small, frame=single, breaklines=true, caption={JSONL schema used to represent each entry in the dataset.}, label={lst:jsonl_schema}]
{
  "base_case_id": "<numeric case id>",
  "variation_id": "<incremental id>",
  "migration_category": "<crypto family>",
  "prequantum_algorithm": "<classical algorithm>",
  "postquantum_algorithm": "<post-quantum algorithm>",
  "prequantum_code": "<Python code>",
  "postquantum_code": "<Python code>",
  "description": "<short human summary>"
}
\end{lstlisting}

\paragraph{Dataset generation.}
Dataset generation (see 1.C in Fig.~\ref{fig:dataset_method}) followed a two-stage process combining manual seed design and controlled LLM-assisted expansion, followed by human review and filtering. First, a small set of base cases was written entirely by hand. These seed examples were designed to capture core migration patterns and to provide high-quality reference pairs from which broader variations could later be derived. Each manually created pair already satisfied the main design constraints of the dataset: semantic equivalence between the pre- and post-quantum versions, explicit forward and inverse operations, and executable code structure.

Second, the dataset was expanded through controlled interactions with the generative model \texttt{o4-mini-high}. Since code-oriented LLMs typically show stronger performance in English \cite{li2024quantifying,harigai2025response}, all prompts and generated outputs were produced in English. The prompt context included the manually created base cases, the target JSONL schema, naming constraints, and a case guide describing the canonical cryptographic mappings to be followed. To reduce stylistic contamination across families, generation was organized as one dedicated conversation per category.

Listing~\ref{lst:dataset-entry-examples} shows two compact examples of generated dataset entries. Only the metadata, function headers and cryptographically relevant operations are shown; non-essential boilerplate is omitted for readability. The first example illustrates a digital-signature migration from RSA to Dilithium5, while the second shows a key-establishment migration from Diffie--Hellman to Kyber1024. These examples highlight that the generated entries are not isolated primitive substitutions: each pair preserves the operational structure needed for later functional validation.

\begin{lstlisting}[
basicstyle=\ttfamily\scriptsize,
frame=single,
breaklines=true,
caption={Compact examples of generated dataset entries. Only metadata, function headers and cryptographically relevant operations are shown.},
label={lst:dataset-entry-examples}
]
# Example 1: digital signature migration
{
  "base_case_id": "1",
  "variation_id": "1.1",
  "migration_category": "digital signature",
  "prequantum_algorithm": "RSA",
  "postquantum_algorithm": "Dilithium5",

  "prequantum_code": "
    from cryptography.hazmat.primitives.asymmetric import rsa, padding
    from cryptography.hazmat.primitives import hashes

    def generate_keys():
        private_key = rsa.generate_private_key(
            public_exponent=65537,
            key_size=2048
        )
        return private_key, private_key.public_key()

    def sign(private_key, message: bytes) -> bytes:
        return private_key.sign(
            message,
            padding.PSS(
                mgf=padding.MGF1(hashes.SHA256()),
                salt_length=padding.PSS.MAX_LENGTH
            ),
            hashes.SHA256()
        )

    def verify(public_key, message: bytes, signature: bytes) -> bool:
        ...
  ",

  "postquantum_code": "
    import oqs
    import ctypes as ct

    ALGORITHM = 'Dilithium5'

    def generate_keys():
        with oqs.Signature(ALGORITHM) as sig:
            public_key = sig.generate_keypair()
            secret_key = sig.export_secret_key()
        return secret_key, public_key

    def sign(secret_key: bytes, message: bytes) -> bytes:
        clean_key = secret_key.rstrip(b'\\x00')
        with oqs.Signature(ALGORITHM) as sig:
            sig.secret_key = ct.create_string_buffer(
                clean_key, len(clean_key)
            )
            return sig.sign(message)

    def verify(public_key: bytes, message: bytes, signature: bytes) -> bool:
        ...
  "
}

# Example 2: key-establishment migration
{
  "base_case_id": "2",
  "variation_id": "2.1",
  "migration_category": "key exchange",
  "prequantum_algorithm": "Diffie-Hellman",
  "postquantum_algorithm": "Kyber1024",

  "prequantum_code": "
    from cryptography.hazmat.primitives.asymmetric import dh
    from cryptography.hazmat.primitives.kdf.hkdf import HKDF
    from cryptography.hazmat.primitives import hashes

    parameters = dh.generate_parameters(generator=2, key_size=2048)

    def generate_keys():
        private_key = parameters.generate_private_key()
        return private_key, private_key.public_key()

    def derive_shared_secret(private_key, peer_public_key):
        shared_key = private_key.exchange(peer_public_key)
        return HKDF(
            algorithm=hashes.SHA256(),
            length=32,
            salt=None,
            info=b'handshake data'
        ).derive(shared_key)
  ",

  "postquantum_code": "
    import oqs

    ALGORITHM = 'Kyber1024'

    def generate_keys():
        with oqs.KeyEncapsulation(ALGORITHM) as kem:
            public_key = kem.generate_keypair()
            secret_key = kem.export_secret_key()
        return secret_key, public_key

    def encapsulate_shared_secret(peer_public_key):
        with oqs.KeyEncapsulation(ALGORITHM) as kem:
            ciphertext, shared_secret = kem.encap_secret(peer_public_key)
        return ciphertext, shared_secret

    def decapsulate_shared_secret(ciphertext, secret_key):
        ...
  "
}
\end{lstlisting}

Expansion was carried out iteratively in batches of ten variations. After each batch, automatically detectable issues such as malformed JSONL structure or syntax errors were filtered through the testing workflow, and each accepted candidate was also reviewed manually to ensure that the code was semantically meaningful and complied with the migration requirements. This makes the generation process explicitly human-in-the-loop, rather than fully automatic, which is appropriate for a dataset intended to serve as a high-quality benchmark \cite{wu2022survey}. Between batches, reflective prompting was used to identify missing patterns, underrepresented structures, or inconsistencies in the corpus, following the general intuition of iterative self-refinement strategies \cite{madaan2023self,praas2023self}.

The expansion process also aimed to capture structural diversity beyond direct primitive substitution. As a result, the corpus includes heterogeneous programming patterns such as asynchronous execution, context managers, streaming I/O, class-based wrappers, and function-oriented implementations. This variability is deliberate: it exposes the future fine-tuned models to realistic syntactic and structural differences, reducing the risk that migration is learned as a purely lexical replacement task.

\paragraph{Dataset validation.}
Dataset validation (see 1.D in Fig.~\ref{fig:dataset_method}) was integrated directly into dataset construction. The same automated verification logic that later supports model evaluation was first used to verify the correctness of the dataset pairs themselves. This gives the validation stage a dual methodological role: it guarantees the quality of the benchmark during corpus creation and later serves as the evaluation oracle for model-generated migrations.

In practice, validation combines two complementary checks. First, syntactic validity is assessed during controlled execution inside the testing harness: malformed snippets fail immediately when they are loaded and executed. Second, dynamic functional tests verify that both the pre-quantum and post-quantum implementations satisfy the expected property of the corresponding cryptographic task. These properties depend on the category and include, for example, successful encrypt/decrypt round-trips, valid sign/verify flows, correct encapsulation/decapsulation behavior, or correct tag generation and verification in MAC settings.

The verification workflow is organized by category. For each cryptographic family, a dedicated test file implements shared validation logic for the common cases and specialized subtests for variations with more specific calling conventions or code structures. A variation is considered correct and added to the final dataset only when the corresponding functional property is satisfied independently by both the pre-quantum and post-quantum implementations. The test harness prints structured JSON summaries reporting the total number of evaluated cases, the number of successful cases, and the identifiers of failed instances. These summaries provide the basis for the later quantitative evaluation of fine-tuned models.

To make the validation logic concrete, Listings~\ref{lst:signature-validation-example} and~\ref{lst:kex-validation-example} show two representative examples from the dataset. The first illustrates a digital-signature migration from RSA to Dilithium5, where correctness is defined as successful signing and verification in both the pre-quantum and post-quantum branches. The second illustrates a key-establishment migration from Diffie--Hellman to Kyber1024, where correctness is defined by shared-secret agreement in the classical branch and encapsulation/decapsulation agreement in the post-quantum branch.

\begin{lstlisting}[
basicstyle=\ttfamily\scriptsize,
frame=single,
breaklines=true,
caption={Representative validation pattern for a digital-signature migration from RSA to Dilithium5.},
label={lst:signature-validation-example}
]
# Dataset entry: RSA -> Dilithium5
# Functional property: sign/verify succeeds in both branches.

TEST_MESSAGE = b"Test digital signature message."

def validate_signature_migration(variation, mapping):
    pre_ns, post_ns = {}, {}

    # Load pre-quantum and post-quantum snippets.
    exec(variation["prequantum_code"], pre_ns)
    exec(variation["postquantum_code"], post_ns)

    # --- Pre-quantum branch: RSA sign/verify ---
    pre_keygen = pre_ns[mapping["generate_keys"]]
    pre_sign = pre_ns[mapping["pre_sign"]]
    pre_verify = pre_ns[mapping["pre_verify"]]

    pre_priv, pre_pub = pre_keygen()
    pre_sig = pre_sign(pre_priv, TEST_MESSAGE)

    if not pre_verify(pre_pub, TEST_MESSAGE, pre_sig):
        raise ValueError("Pre-quantum verification failed.")

    # --- Post-quantum branch: Dilithium5 sign/verify ---
    post_keygen = post_ns[mapping["generate_keys"]]
    post_sign = post_ns[mapping["post_encrypt"]]
    post_verify = post_ns[mapping["post_decrypt"]]

    post_priv, post_pub = post_keygen()
    post_sig = post_sign(post_priv, TEST_MESSAGE)

    if not post_verify(post_pub, TEST_MESSAGE, post_sig):
        raise ValueError("Post-quantum verification failed.")

    return True
\end{lstlisting}

\begin{lstlisting}[
basicstyle=\ttfamily\scriptsize,
frame=single,
breaklines=true,
caption={Representative validation pattern for a key-establishment migration from Diffie--Hellman to Kyber1024.},
label={lst:kex-validation-example}
]
# Dataset entry: Diffie-Hellman -> Kyber1024
# Functional property:
# - classical branch: both parties derive the same secret
# - PQ branch: encapsulation and decapsulation produce the same secret

def validate_key_exchange_migration(variation, mapping):
    pre_ns, post_ns = {}, {}

    # Load pre-quantum and post-quantum snippets.
    exec(variation["prequantum_code"], pre_ns)
    exec(variation["postquantum_code"], post_ns)

    # --- Pre-quantum branch: Diffie-Hellman shared secret ---
    pre_generate = pre_ns[mapping["pre_generate_keys"]]
    pre_derive = pre_ns[mapping["pre_derive_shared_secret"]]

    pre_priv_a, pre_pub_a = pre_generate()
    pre_priv_b, pre_pub_b = pre_generate()

    pre_shared_a = pre_derive(pre_priv_a, pre_pub_b)
    pre_shared_b = pre_derive(pre_priv_b, pre_pub_a)

    if pre_shared_a != pre_shared_b:
        raise ValueError("Pre-quantum key agreement failed.")

    # --- Post-quantum branch: Kyber1024 KEM flow ---
    post_generate = post_ns[mapping["post_generate_keys"]]
    post_encap = post_ns[mapping["post_encapsulate_shared_secret"]]
    post_decap = post_ns[mapping["post_decapsulate_shared_secret"]]

    post_secret_key, post_public_key = post_generate()

    ciphertext, shared_enc = post_encap(post_public_key)
    shared_dec = post_decap(ciphertext, post_secret_key)

    if shared_enc != shared_dec:
        raise ValueError("Post-quantum KEM validation failed.")

    return True
\end{lstlisting}

\section{LLM training and testing}
\label{sec:LLM_general}

\begin{figure}
    \centering
    \includegraphics[width=1\linewidth]{ 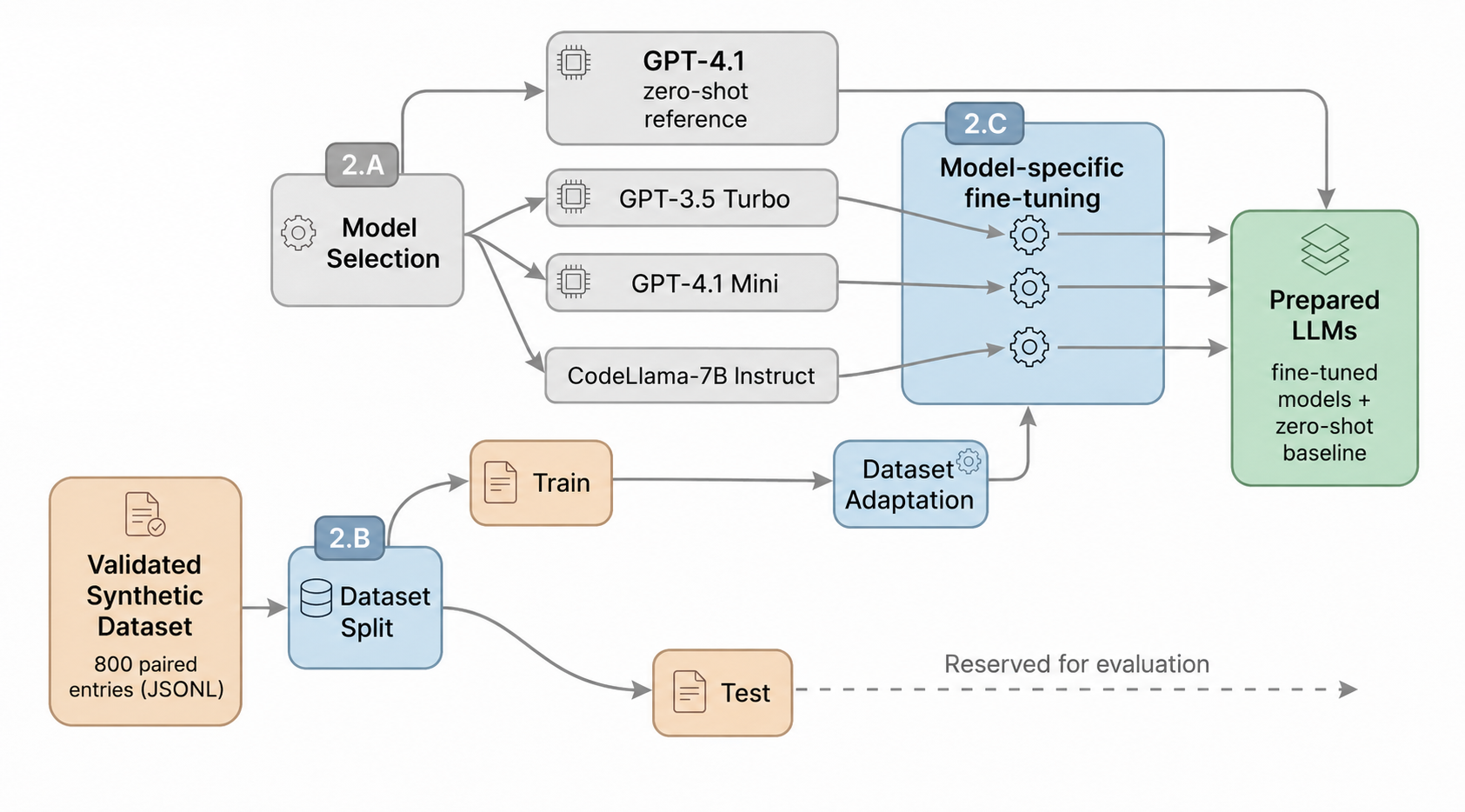}
    \caption{\textit{LLM Preparation Phase}. The validated synthetic dataset is split into train and validation subsets. Only the train split is used for dataset adaptation and model-specific fine-tuning, while the validation split is reserved for the evaluation phase. GPT-3.5 Turbo, GPT-4.1 Mini, and CodeLlama-7B Instruct undergo fine-tuning, whereas GPT-4.1 is included as a zero-shot reference and bypasses the preparation stage.}
    \label{fig:model_preparation}
\end{figure}

\paragraph{Model selection.}
\label{sec:model-selection}
Selecting appropriate models was essential to evaluate the feasibility of automated cryptographic migration (see 2.A in Fig.~\ref{fig:model_preparation}). The study sought to include both proprietary and open-source architectures to ensure that the proposed methodology could be reproduced across computational and licensing constraints. The selection prioritized diversity in scale, accessibility, and architecture while maintaining compatibility with modern instruction-tuning frameworks.

The selection of models followed a staged process aimed at covering a representative spectrum of instruction-tuned code LLMs available in 2025: state-of-the-art proprietary systems, widely adopted mid-range models, and open-source architectures suitable for local experimentation. This diversity is essential in cryptographic migration, where the goal is not only to assess raw performance, but also reproducibility, cost-efficiency, and feasibility under constrained hardware and budgets.

From the proprietary side, OpenAI’s models stand out for their consistent leadership in coding benchmarks such as HumanEval+ and MBPP+ \cite{openai2025gpt41}. Their API provides a mature fine-tuning pipeline, deterministic message formatting (ChatML), extended context windows up to one million tokens, and precise per-token usage accounting \cite{openai2025pricing,openai2025gpt41docs}. Such properties make them strong candidates for controlled scientific evaluation, where traceability and repeatability are paramount.

In contrast, open-source alternatives play a complementary role: they allow full control over the training environment, zero API cost, complete reproducibility, and the ability to inspect intermediate representations. Among them, CodeLlama stands out as Meta’s code-specialized LLaMA variant, trained on 500B tokens of source code and evaluated extensively on code-generation benchmarks \cite{roziere2023code}. When combined with low-rank adaptation and 4-bit quantization, CodeLlama-7B can be fine-tuned on commodity GPUs using QLoRA \cite{dettmers2023qlora}, making it an accessible option for academic laboratories and low-resource research environments.

Next, we list the models selected and the rationale behind the selection:
\begin{itemize}
    \item \textbf{GPT-4.1}.  
    Included as the zero-shot reference model in the original experimental protocol. It represents the strongest general-purpose model publicly available through API at the time of the study, with top-tier coding accuracy and a 1M-token context window \cite{openai2025gpt41docs}. Using GPT-4.1 without prior fine-tuning establishes a realistic ceiling for what can be achieved without domain-specific adaptation.

    \item \textbf{GPT-3.5-turbo}.  
    By 2024–2025, GPT-3.5 remained one of the most widely deployed LLMs in both industry and academia due to its stability, mature documentation and extremely low operational cost \cite{openai2025pricing}. Its widespread adoption is reflected in several software-engineering studies that rely on GPT-3.5-Turbo as a practical baseline, including work on automatic refactoring where the model achieves high accuracy in few-shot settings \cite{shirafuji2023refactoring}. Including GPT-3.5-turbo fine-tuning therefore enables direct comparison with a model still present in a large fraction of real-world systems and commonly used in academic benchmarks.

    \item \textbf{GPT-4.1-mini}.  
    Chosen as the primary fine-tuning target due to its favourable balance between cost and performance. According to OpenAI’s public evaluations, GPT-4.1-mini matches or exceeds GPT-4o in several reasoning and coding tasks at a fraction of the cost \cite{openai2025gpt41}. Its training and inference prices (up to 80\% lower than GPT-4.1) make large-scale experiments feasible within a restricted budget \cite{openai2025pricing}. Critically, it also inherits the full 1M-token context window, enabling long examples to be included without segmentation.

    \item \textbf{CodeLlama-7B-Instruct}.  
    Selected as the open-source baseline to test the replicability of the pipeline without proprietary APIs. CodeLlama is designed specifically for code and performs competitively on benchmarks relative to much larger open models \cite{roziere2023code}. Using QLoRA \cite{dettmers2023qlora}, the 7B variant can be fine-tuned on GPUs with 8–16 GB of VRAM, demonstrating that cryptographic migration can be studied even in resource-limited academic settings. Its inclusion also enables direct comparison between closed and open ecosystems.
\end{itemize}

In summary, the final selection spans three axes crucial for this study: performance ceiling, industry relevance, and open-source feasibility (see Table~\ref{tab:model-selection-summary}). This combination ensures that the evaluation captures not only state-of-the-art behavior but also the practical trade-offs faced by organizations that must deploy PQC migration tools under heterogeneous constraints.

\begin{table*}[t]
\centering
\small
\caption{Summary of selected models and rationale.}
\label{tab:model-selection-summary}
\begin{tabular}{p{2.4cm} p{2cm} p{4.4cm} p{2cm} p{4.7cm}}
\hline
\textbf{Model} & \textbf{Ecosystem} & \textbf{Role in study} & \textbf{Fine-tuned} & \textbf{Notes / constraints} \\
\hline
GPT-4.1 & Prop. (API) & Zero-shot upper-bound reference & No & 1M-token context; strong coding baseline \\
GPT-3.5-turbo & Prop. (API) & Widely deployed, low-cost baseline & Yes & Mature documentation; low operational cost \\
GPT-4.1-mini & Prop. (API) & Primary fine-tuning target (cost/perf) & Yes & 1M-token context; up to 80\% cheaper than GPT-4.1 \\
CodeLlama-7B-Instruct & OSS (local) & Open baseline for pipeline replicability & Yes (QLoRA) & 4-bit QLoRA; commodity GPUs (8--16 GB VRAM) \\
\hline
\end{tabular}
\end{table*}

\paragraph{Experimental setting.}
\label{sec:model_configuration}
To ensure methodological parity across all evaluated models, every experiment was conducted under a uniform configuration: an 8k-token context window, identical normalization and whitespace handling, and a shared tokenization layer based on \texttt{tiktoken}. Although \texttt{tiktoken} is natively optimized for OpenAI models, its systematic use as a measurement layer allowed fair comparison of token consumption across all architectures, including the open-source models. This alignment follows standard evaluation practices in LLM benchmarking \cite{yang2023harnessingpowerllmspractice,liu2024understanding}.

A crucial consideration in cross-model evaluation is that different model families rely on different conversational abstractions. The OpenAI models (GPT-4.1, GPT-3.5-turbo, GPT-4.1-mini) operate using a role-based chat interface (system, user, assistant), following the ChatML format. This interface is natively supported by the API and has been shown to stabilize instruction following \cite{yang2023harnessingpowerllmspractice}. Consequently, the structured prompt used for these models mirrors the internal message hierarchy (see Fig~\ref{fig:chatml_prompt_style}).

\begin{figure}
\centering
\begin{tcolorbox}[
    colback=gray!5, colframe=gray!40, width=0.9\linewidth,
    fonttitle=\bfseries]
\textbf{SYSTEM:} \\
You are a code assistant that migrates pre-quantum cryptographic code to post-quantum code.\\

\textbf{USER:} \\
MIGRATION CATEGORY: \texttt{<category>} \\
PRE-QUANTUM CODE: \texttt{<python\_snippet>}
\end{tcolorbox}

\caption{ChatML-style prompt used for OpenAI models}
\label{fig:chatml_prompt_style}
\end{figure}

This two-message structure was kept constant for all OpenAI models, regardless of whether they were used in zero-shot mode (GPT-4.1) or fine-tuned (GPT-3.5-turbo, GPT-4.1-mini). The uniformity of this format isolates the effect of fine-tuning and avoids confounding variables arising from prompt engineering.

In contrast, decoder-only open-source models such as CodeLlama-7B-Instruct do not implement a role-based chat protocol. Instead, they use an instruction boundary format built around the \texttt{[INST] ... [/INST]} markers, which is consistent with the LLaMA instruction-tuning methodology reported in the literature \cite{touvron2023llama}. For these models, the same semantic content (category + pre-quantum code) must be serialized into a single completion prompt (see Fig~\ref{fig:codellama_prompt_style}).

\begin{figure}
\centering
\begin{tcolorbox}[
    colback=gray!5, colframe=gray!40, width=0.9\linewidth,
    fonttitle=\bfseries]
\textbf{[INST]} \\
You are a code assistant that migrates pre-quantum cryptographic code to post-quantum code. \\
MIGRATION CATEGORY: \texttt{<category>} \\
PRE-QUANTUM CODE: \texttt{<python\_snippet>}
\textbf{[/INST]}
\end{tcolorbox}

\caption{Instruction-style prompt for CodeLlama models}
\label{fig:codellama_prompt_style}
\end{figure}

The model then learns to produce the post-quantum code directly after the \texttt{[/INST]} token. This {completion-only} training scheme is standard in instruction-tuned LLaMA-family models, where the text following the closing marker constitutes the supervised target during fine-tuning \cite{touvron2023llama}. This setup aligns with the expectations of the TRL \texttt{SFTTrainer}, which implements supervised instruction tuning by treating only the post-prompt segment as the label sequence, a method widely adopted in open-source adaptation pipelines \cite{dettmers2023qlora}.

Despite these interface differences, both prompting strategies encode the same information: (1) the migration task definition, (2) the cryptographic primitive category, and (3) the input code to transform. Maintaining semantic equivalence ensures that the comparison between OpenAI chat models and a locally trained CodeLlama remains meaningful and technically fair, as recommended in cross-family LLM benchmarking \cite{yang2023harnessingpowerllmspractice,liu2024understanding}.

Finally, normalization of prompts into structured templates—either ChatML or \texttt{[INST]}—ensures that model behaviour reflects differences in architecture and training, not inconsistencies in how instructions are presented. This consistency is essential for isolating model-specific effects in downstream analyses such as static similarity, functional correctness, and error patterns emerging during cryptographic migration.

OpenAI-hosted models were fine-tuned through the official API pipeline, while the open-source configuration relied on the \texttt{HuggingFace Transformers} ecosystem with quantized fine-tuning via \texttt{bitsandbytes} and \texttt{QLoRA}. This setup reduced the memory footprint of \texttt{CodeLlama-7B} to 4-bit precision, allowing execution on a single high-end GPU without compromising numerical stability. All environments were instrumented with deterministic logging and version control to ensure exact reproducibility.

This configuration established a controlled foundation for the fine-tuning and evaluation phases.

\paragraph{Dataset split for training and testing.}
A crucial step in preparing the corpus for model training was the construction of a clean, balanced and statistically meaningful dataset split (see 2.B in Fig.~\ref{fig:model_preparation}). Following established practices in supervised LLM training and dataset curation \cite{de2024nos_corpusnos,wu2022survey}, the 800 validated examples were divided into 720 training samples and 80 held-out validation samples.

The partition was generated as a random stratified holdout split. Stratification was performed by migration category, ensuring that every cryptographic family was represented in both the training and validation subsets, while randomness avoided manually selecting favourable validation examples. For each core cryptographic category--hashing, message authentication codes (MAC), digital signatures, key exchange, symmetric encryption, and authenticated encryption--exactly 90 samples were allocated to the training split and 10 samples to the validation split. This ensured that every primitive was equally represented during optimization and evaluation, avoiding biases caused by uneven category frequencies.

Combined cases, i.e., snippets involving two different cryptographic primitive families, were handled separately due to their higher structural complexity. These entries were distributed so that both the training and validation subsets preserved a comparable mix of primitive-pair interactions. This design choice reduces the risk of overfitting to single-primitive patterns while still exposing the model to realistic multi-primitive interactions, which are common in production systems \cite{wu2022survey}.

Stratification is particularly important in cryptographic migration tasks, where different primitives exhibit substantially different code structures, dependency patterns and verification requirements. By preserving proportional representation across both simple and combined categories, the resulting splits enable reliable generalization analysis and fair cross-model comparison.

A full k-fold cross-validation protocol was considered but not adopted as the main evaluation strategy. In conventional machine-learning settings, k-fold validation is often feasible because training can be repeated many times at moderate cost. In this study, however, each fold would require launching an independent fine-tuning run for each model under evaluation. For hosted models, this implies creating a new training file and a separate fine-tuning job for every fold through the API; for local models, it implies repeating the full GPU fine-tuning process. Consequently, a 5- or 10-fold protocol would multiply training time, inference time and cost by the number of folds, making it disproportionate for an 800-example exploratory study. Moreover, the purpose of this work is not to estimate a single production-grade generalization score, but to compare model behaviour under a controlled, reproducible validation protocol.

However, to reduce the risk that the reported results depend on a single favourable split, an additional robustness check was performed using a second random stratified holdout partition. This alternative split used the same dataset size and proportions as the original experiment--720 training examples and 80 validation examples--but changed the random seed. The same \texttt{GPT-4.1-mini} fine-tuning and evaluation procedure was then repeated on this new partition. This repeated stratified holdout does not replace a full k-fold protocol, but it provides additional evidence that the observed performance is not an artefact of a uniquely favourable train/validation split.

Table~\ref{tab:dataset_split_overview} provides a compact overview of the dataset composition, while the complete distribution of cryptographic primitives across training and validation splits is reported in ~\ref{app:dataset} for reproducibility and transparency.

\begin{table}[t]
\centering
\caption{Summary of dataset composition by cryptographic category.}
\label{tab:dataset_split_overview}
\small
\begin{tabular}{lcc}
\toprule
\textbf{Category} & \textbf{Train} & \textbf{Validation} \\
\midrule
Hash functions & 90 & 10 \\
MAC & 90 & 10 \\
Digital signatures & 90 & 10 \\
Key exchange / KEM & 90 & 10 \\
Symmetric encryption & 90 & 10 \\
Authenticated encryption (AEAD) & 90 & 10 \\
Combined primitives & 180 & 20 \\
\midrule
\textbf{Total} & \textbf{720} & \textbf{80} \\
\bottomrule
\end{tabular}
\end{table}

\paragraph{LLM training and testing}
\label{sec:training}

Each model selected was trained on the same dataset and later applied to unseen code fragments to assess its capacity for reliable cryptographic migration (see 2.C in Fig.~\ref{fig:model_preparation}). {Notice that while the model training is the last step in model preparation (Fig.~\ref{fig:model_preparation} details this phase), Fig~\ref{fig:evaluation} provides an overview of the  model testing (evaluation) pipeline.} 

After model selection, the training phase focused on adapting the dataset and training workflow to each model family. All preprocessing, dataset conversion, prompt formatting, evaluation scripts, and local fine-tuning code were implemented in Python. For OpenAI-hosted models, fine-tuning was executed through the official API, while the local CodeLlama-7B-Instruct configuration relied on the HuggingFace \texttt{transformers} ecosystem and \texttt{trl} for supervised fine-tuning. This distinction allowed both hosted and local models to be evaluated under a common experimental protocol while respecting the different interfaces required by each architecture.

The fine-tuning process was supervised and instruction-based, aligning the model’s output generation with the expected format of pre- and post-quantum code pairs. Each training instance consisted of a dialogue-style record following the ChatML convention: a system message defining the migration context, a user message containing the pre-quantum code and its cryptographic category, and an assistant message providing the corresponding post-quantum version. This conversational structure preserved alignment between fine-tuning and inference, enabling the model to generalize its reasoning to new unseen examples.

Fine-tuning was carried out over approximately one million tokens using a learning rate of $2\times10^{-4}$, batch size of 4, and six epochs. For local experiments, \texttt{CodeLlama-7B} was trained using \texttt{QLoRA}, which allowed efficient adaptation on a single NVIDIA A100 GPU. Hosted OpenAI models were trained through the fine-tuning API under equivalent hyperparameter schedules. Validation loss was monitored after each epoch to detect overfitting and to confirm stable convergence.

Once trained, the models were evaluated on the held-out validation split, e.g., code samples not seen during fine-tuning (see Fig.~\ref{fig:evaluation}). The migration pipeline emulated a realistic workflow: source files were loaded, automatically segmented when exceeding the context limit, and enriched with structured prompts specifying their cryptographic category. The model then generated the migrated code, which was reconstructed into full executable files. No automatic post-processing was applied at this stage to preserve the raw output for controlled analysis. {Details on the approach taken to evaluate the models' performance when asked to migrate the code fragments of the validation dataset are described next in Section \ref{sec:evaluation}.}

\section{Automated evaluation of code fragments migration}
\label{sec:evaluation}

\begin{figure}
    \centering
    \includegraphics[width=1\linewidth]{ 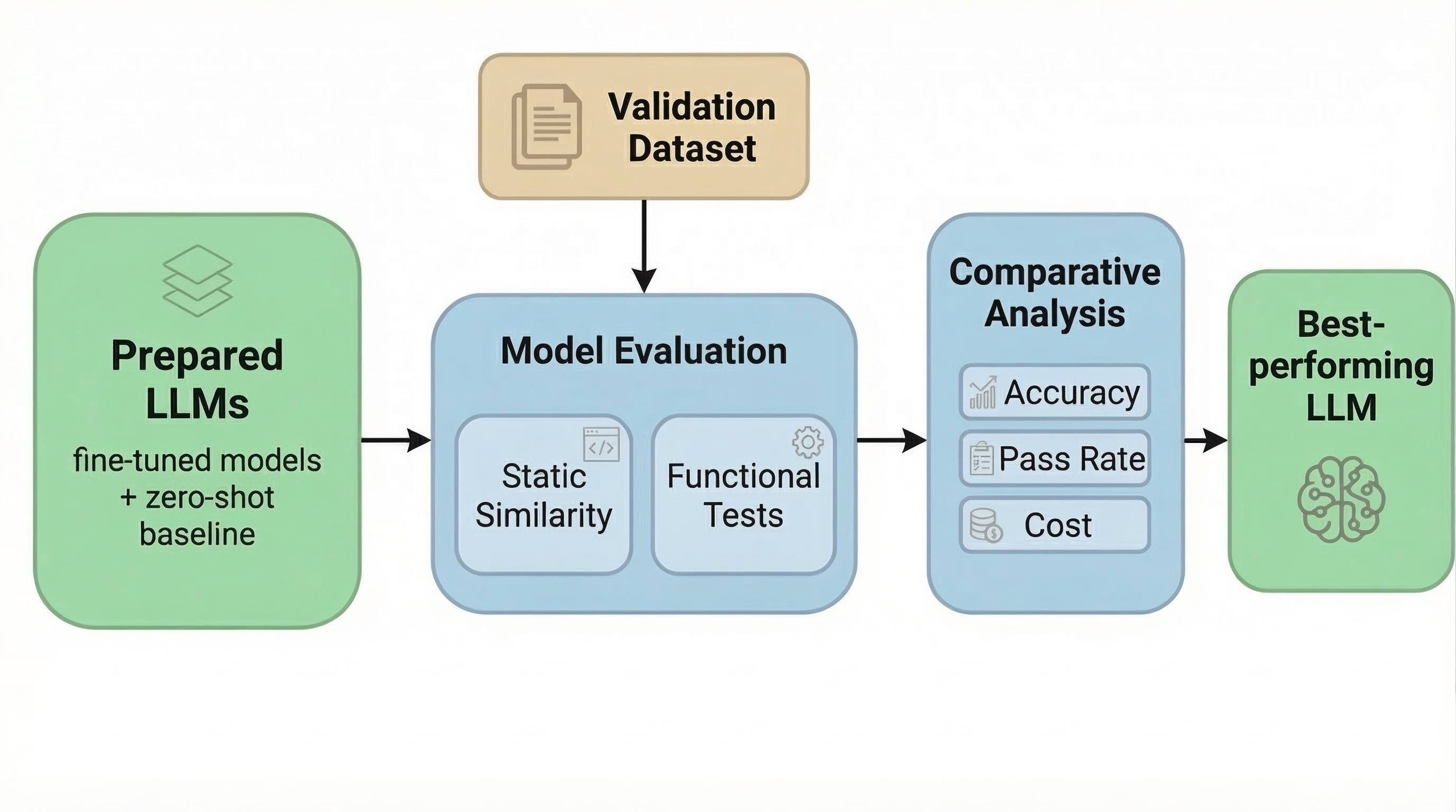}
    \caption{\textit{LLM Evaluation Phase}. The prepared LLMs are evaluated on the validation dataset using both static similarity metrics and automated functional tests. Their results are then compared through a multi-criteria analysis, leading to the selection of the best-performing LLM for the final real-world validation stage.}
    \label{fig:evaluation}
\end{figure}

Evaluating the effectiveness of large language models in cryptographic migration requires more than syntactic comparison between generated and reference code. Functional correctness, reproducibility, and error traceability are equally critical, as cryptographic transformations must preserve both algorithmic behavior and structural integrity. The evaluation framework developed for this study was therefore designed to measure multiple dimensions of model performance, combining static and dynamic verification within a unified testing pipeline. Fig.~\ref{fig:evaluation} outlines the end-to-end procedure used to evaluate the models.

Recall that per dataset design, each dataset entry is paired not only with pre-quantum and post-quantum code, but also with a test designed during dataset construction to validate the expected functionality of that specific snippet (e.g., encryption/decryption round-trips, signing/verification). The post-quantum code in each dataset entry can be used to assess statically the output of the LLM being evaluated. The per-entry test oracle enables functional (dynamical) checking at scale and is used both to validate the dataset itself and to evaluate model outputs on unseen samples.

Model evaluation is performed through a Python-based test harness that dispatches each validation entry to the appropriate category-specific validator and executes the corresponding test on the generated (or migrated) code. The harness outputs structured JSON summaries reporting the total number of evaluated cases, aggregated success metrics, and the identifiers of failed instances for each model.

\paragraph{Metrics.} 
As already introduced, two complementary evaluation dimensions were considered. The first was \textbf{static similarity}, assessing how closely the generated code matched the expected structure. This metric was computed using Python’s \texttt{SequenceMatcher}, producing a normalized lexical-structural similarity score between 0 and 1. It provided an initial approximation of whether the model preserved code structure, naming conventions, and syntax across transformations. However, similarity alone was insufficient to judge correctness, as superficial lexical overlap can mask deeper semantic inconsistencies.

To address this, a second dimension, \textbf{dynamic functional correctness}, was incorporated using the test oracle present in each dataset entry. Each generated script by the LLM was executed in an isolated environment, where the tests validated its functional behaviour. Depending on the cryptographic category, these tests checked whether the outputs of encryption/decryption, sign/verify, encapsulate/decapsulate, or MAC generation/verification cycles were consistent with the expected reference values. This layer of testing captured runtime errors, incorrect parameter handling, or key mismanagement that might pass unnoticed in purely static analysis.

The two metrics were combined to produce a comprehensive performance profile. Static similarity served as a proxy for structural fidelity, while dynamic verification measured true functional equivalence. The dual metric enabled fine-grained comparison between models and configurations, revealing whether apparent textual precision translated into real cryptographic correctness.

Additionally, \textbf{token usage} was logged for every inference call, enabling cost-efficiency analysis. For hosted models, API usage statistics were aggregated automatically, while local runs collected system-level performance metrics such as GPU memory and throughput. These auxiliary indicators helped evaluate the scalability of the proposed workflow for larger datasets and industrial applications.

\paragraph{Post-processing of hardcoded key sizes before validation.} Error analysis was performed manually to categorize the types of failures observed during dynamic testing. Common issues included incorrect key sizes, uninitialized variables, and misinterpreted data types (e.g., converting byte objects to integers). A particularly recurrent problem involved {character-counting errors in hardcoded keys}. Consistent with prior findings in the literature \cite{fu2024large}, LLMs frequently miscount characters when asked to generate fixed-length cryptographic keys (for example, 32 bytes), or confuse bytes with textual characters. To mitigate these failures, a lightweight post-processing step was introduced to automatically correct the length of hardcoded keys before validation. This simple adjustment significantly reduced false negatives during dynamic testing and improved the functional validity of the generated migrations.

\section{Results}
\label{sec:results}

\begin{table*}[t]
\centering
\caption{Cross-summary of the main metrics.}
\label{tab:comparacion-global}
\small
\begin{tabular}{
    l
    S[table-format=1.4]
    S[table-format=2.0]
    S[table-format=2.1]
    S[table-format=1.2]
    S[table-format=1.2]
}
\toprule
\multicolumn{1}{c}{\textbf{Model}} 
& \multicolumn{1}{c}{\textbf{Mean sim.}} 
& \multicolumn{1}{c}{\textbf{HIGH}} 
& \multicolumn{1}{c}{\textbf{Dyn.\,\%}} 
& \multicolumn{1}{c}{\textbf{Cost [\$]}} 
& \multicolumn{1}{c}{\textbf{Train tokens [M]}} \\ 
\midrule

GPT-4.1 (ZS)                
& 0.3941 
& 3  
& 15.0 
& 0.21 
& 0.00 \\

CodeLlama-7B (FT)\textsuperscript{*} 
& 0.6031 
& 16 
& 61.3 
& 3.16 
& 2.96 \\ 

GPT-3.5-turbo (FT)         
& 0.8669 
& 50 
& 68.8 
& 8.48 
& 1.01 \\

\textbf{GPT-4.1-mini (FT)}  
& \textbf{0.9072} 
& \textbf{56} 
& \textbf{92.5} 
& \textbf{5.39} 
& \textbf{1.04} \\
\bottomrule
\multicolumn{6}{l}{\footnotesize *Values after corrective post-processing.}
\end{tabular}
\end{table*}

The evaluation produced a detailed view of how large language models behave when trained for cryptographic code migration. Across the configurations tested, the results demonstrate that fine-tuning on domain-specific data is not only beneficial but necessary to achieve reliable and verifiable transformations.

\paragraph{Static similarity and dynamic functional correctness.} Four models were evaluated: \texttt{GPT-4.1} in a zero-shot configuration, and three fine-tuned variants: \texttt{GPT-4.1-mini}, \texttt{GPT-3.5-turbo} and \texttt{CodeLlama-7B}. The fine-tuning process for \texttt{GPT-3.5-turbo} followed the same methodology and dataset as \texttt{GPT-4.1-mini}, but delivered slightly inferior results at a significantly higher cost (Table~\ref{tab:comparacion-global}). Therefore, to focus the discussion on the most representative trade-offs, the remainder of this section concentrates on the three most illustrative configurations: zero-shot (\texttt{GPT-4.1}), OpenAI-hosted fine-tuned (\texttt{GPT-4.1-mini}), and open-source locally fine-tuned (\texttt{CodeLlama-7B}).

\begin{figure*}[t]
    \centering
    \begin{subfigure}[t]{0.32\textwidth}
        \centering
        \includegraphics[width=\linewidth]{ 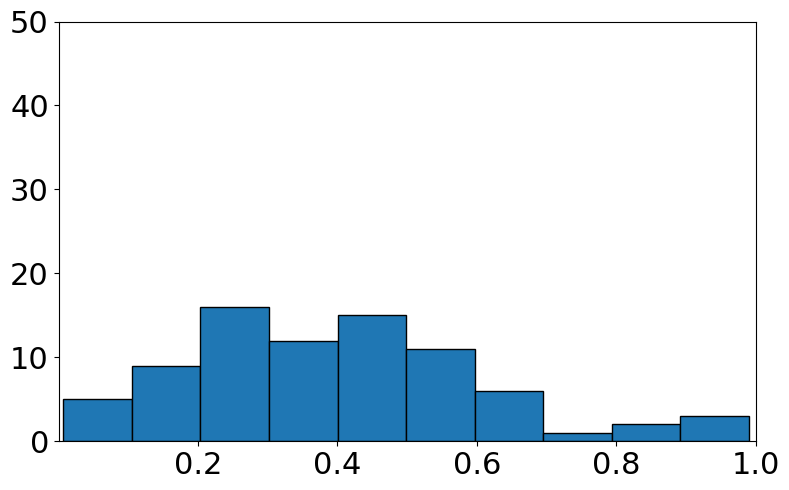}
        \caption{\texttt{GPT-4.1} (zero-shot)}
        \label{fig:stat_gpt41}
    \end{subfigure}
    \hfill
    \begin{subfigure}[t]{0.32\textwidth}
        \centering
        \includegraphics[width=\linewidth]{ 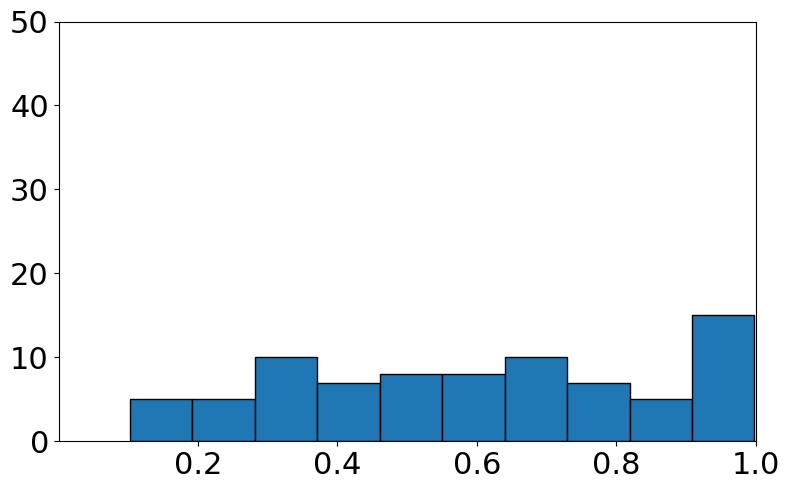}
        \caption{\texttt{CodeLlama-7B} (fine-tuned)}
        \label{fig:stat_codellama}
    \end{subfigure}
    \hfill
    \begin{subfigure}[t]{0.32\textwidth}
        \centering
        \includegraphics[width=\linewidth]{ 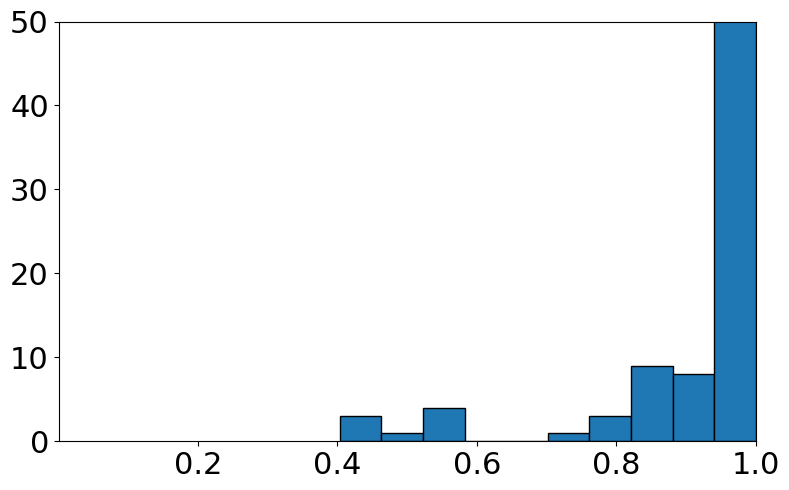}
        \caption{\texttt{GPT-4.1-mini} (fine-tuned)}
        \label{fig:stat_gpt41mini}
    \end{subfigure}

    \caption{Static similarity distributions for pre- to post-quantum code migration. The horizontal axis represents similarity score ranges, while the vertical axis reports the number of code variations falling within each range, reflecting the lexical and structural alignment between original and migrated implementations.}
    \label{fig:static_comparison}
\end{figure*}

\textbf{{GPT-4.1}.} In the zero-shot configuration, general-purpose models such as \texttt{GPT-4.1} displayed strong syntactic fluency but poor semantic precision. As illustrated in Fig.~\ref{fig:stat_gpt41}, the similarity scores are widely spread and concentrated far from perfect overlap with the reference solutions, indicating frequent deviations in the migrated code even when the surrounding structure is preserved.

\begin{figure*}[t]
    \centering
    \begin{subfigure}{0.32\textwidth}
        \centering
        \includegraphics[width=\linewidth]{ 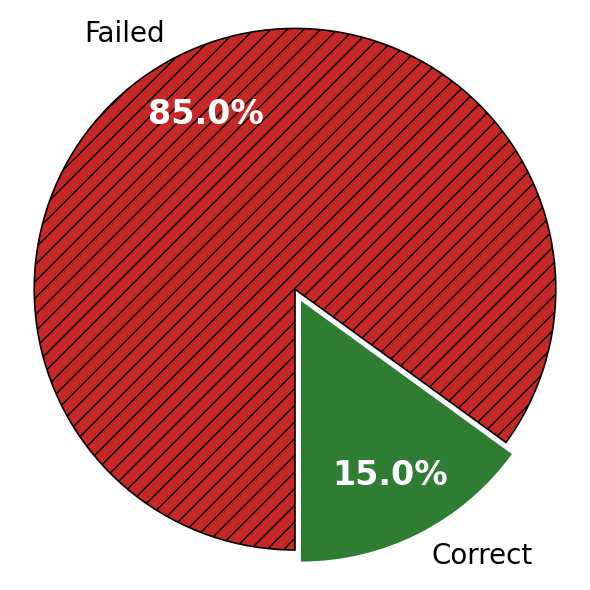}
        \caption{\texttt{GPT-4.1} (zero-shot)}
        \label{fig:dynamic_gpt41}
    \end{subfigure}
    \hfill
    \begin{subfigure}{0.32\textwidth}
        \centering
        \includegraphics[width=\linewidth]{ 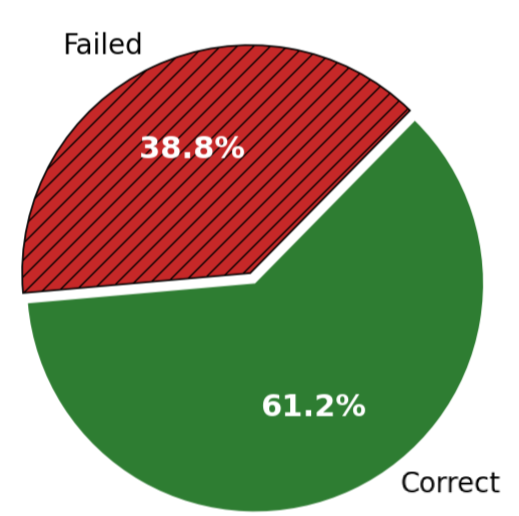}
        \caption{\texttt{CodeLlama-7B} (fine-tuned)}
        \label{fig:dynamic_codellama}
    \end{subfigure}
    \hfill
    \begin{subfigure}{0.32\textwidth}
        \centering
        \includegraphics[width=\linewidth]{ 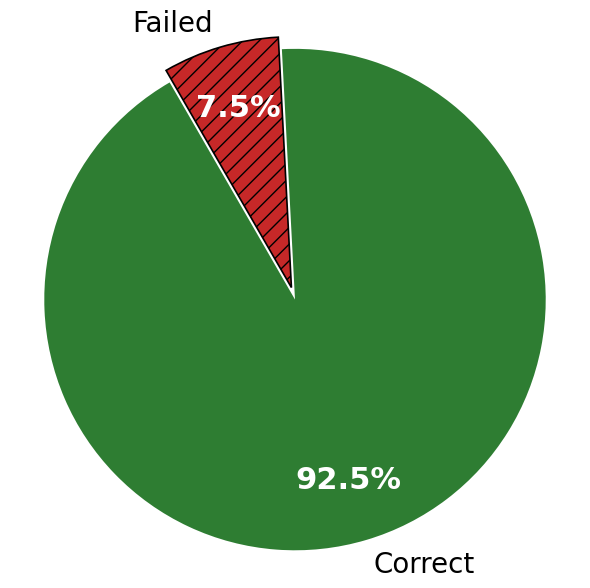}
        \caption{\texttt{GPT-4.1-mini} (fine-tuned)}
        \label{fig:dynamic_gpt41mini}
    \end{subfigure}

    \caption{Dynamic functional correctness across model configurations. 
    Each subfigure shows the distribution of successful and failed executions 
    in the dynamic validation tests.}
    \label{fig:dynamic_comparison}
\end{figure*}

Dynamic validation (Fig.~\ref{fig:dynamic_gpt41}) confirmed that only about 15\,\% of zero-shot \texttt{GPT-4.1} outputs were functionally executable without manual intervention. Typical failures involved incorrect initialization of post-quantum primitives, invalid parameter sizes, or incomplete replacement of legacy calls. This aligns with prior observations that LLMs exhibit high linguistic coherence but limited domain awareness in cryptographic reasoning.

\textbf{{GPT-4.1-mini}.} Fine-tuning led to a substantial improvement across all metrics. For the fine-tuned \texttt{GPT-4.1-mini}, the static distribution in Fig.~\ref{fig:stat_gpt41mini} shifts clearly towards high similarity values, reflecting that most migrations remain very close to the reference solutions at the token and structural level. In dynamic tests (Fig.~\ref{fig:dynamic_gpt41mini}), the model reached a 92.5\,\% success rate, meaning that the vast majority of migrated scripts not only execute successfully (are syntactically valid and runnable under the Python interpreter) but also pass the functional tests (encrypt/decrypt, sign/verify, encaps/decaps) without manual fixes. After evaluating the model performance with a different train split, the results were consistent with the original experiment: mean static similarity increased slightly from 0.907188 to 0.922453, and dynamic functional correctness increased from 74/80 to 78/80 successful tests.

\textbf{{CodeLlama-7B}.} The other fine-tuned baseline shows an intermediate behaviour. As seen in Figs.~\ref{fig:stat_codellama} and \ref{fig:dynamic_codellama}, \texttt{CodeLlama-7B} improves markedly over zero-shot performance but still lags behind \texttt{GPT-4.1-mini} in both static similarity and dynamic correctness. These trends are summarised numerically in Table~\ref{tab:comparacion-global} and visually in Fig.~\ref{fig:static_comparison} and Fig.~\ref{fig:dynamic_comparison}.

\paragraph{Resource consumption.} In terms of cost and efficiency, the fine-tuning process proved economically viable. As reported in Table~\ref{tab:comparacion-global}, the entire experiment for the OpenAI-hosted models, including dataset upload, training and evaluation, required less than \$15 in total API usage. The open-source \texttt{CodeLlama-7B} training completed within six hours on a single NVIDIA A100 GPU using under 24\,GB of memory, confirming the feasibility of reproducing the setup in academic or industrial environments with moderate resources.

\begin{figure}
    \centering
    \includegraphics[width=1\linewidth]{ 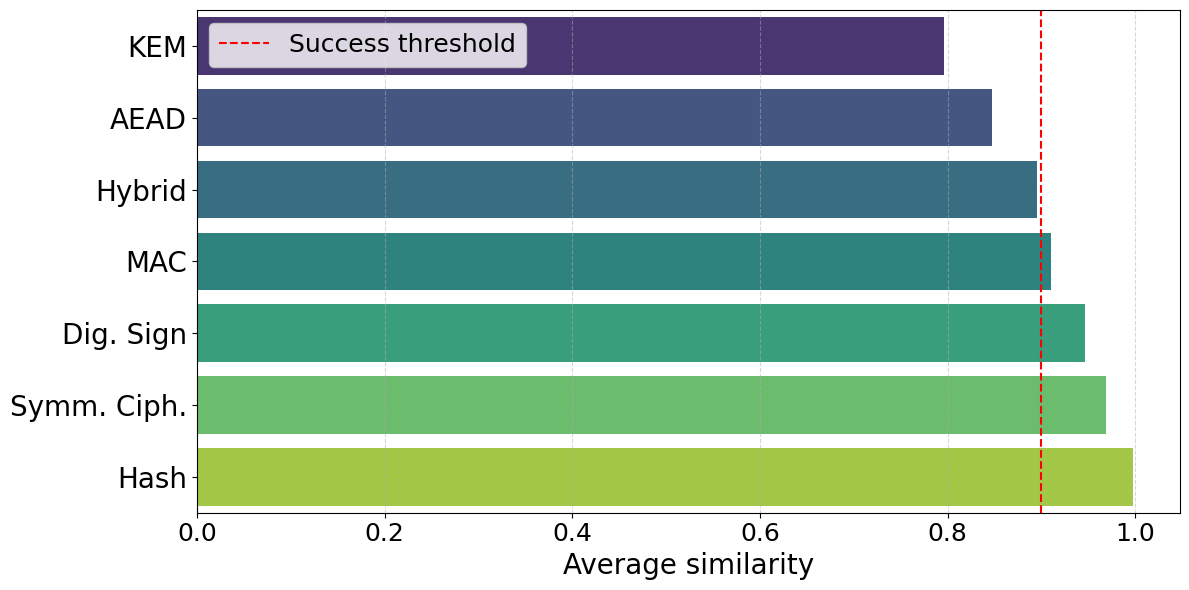}
    \caption{Similarity distribution with predominant category by range (GPT-4.1 mini).}
    \label{fig:gpt41_static_hist}
\end{figure}

\begin{figure}
    \centering
    \includegraphics[width=1\linewidth]{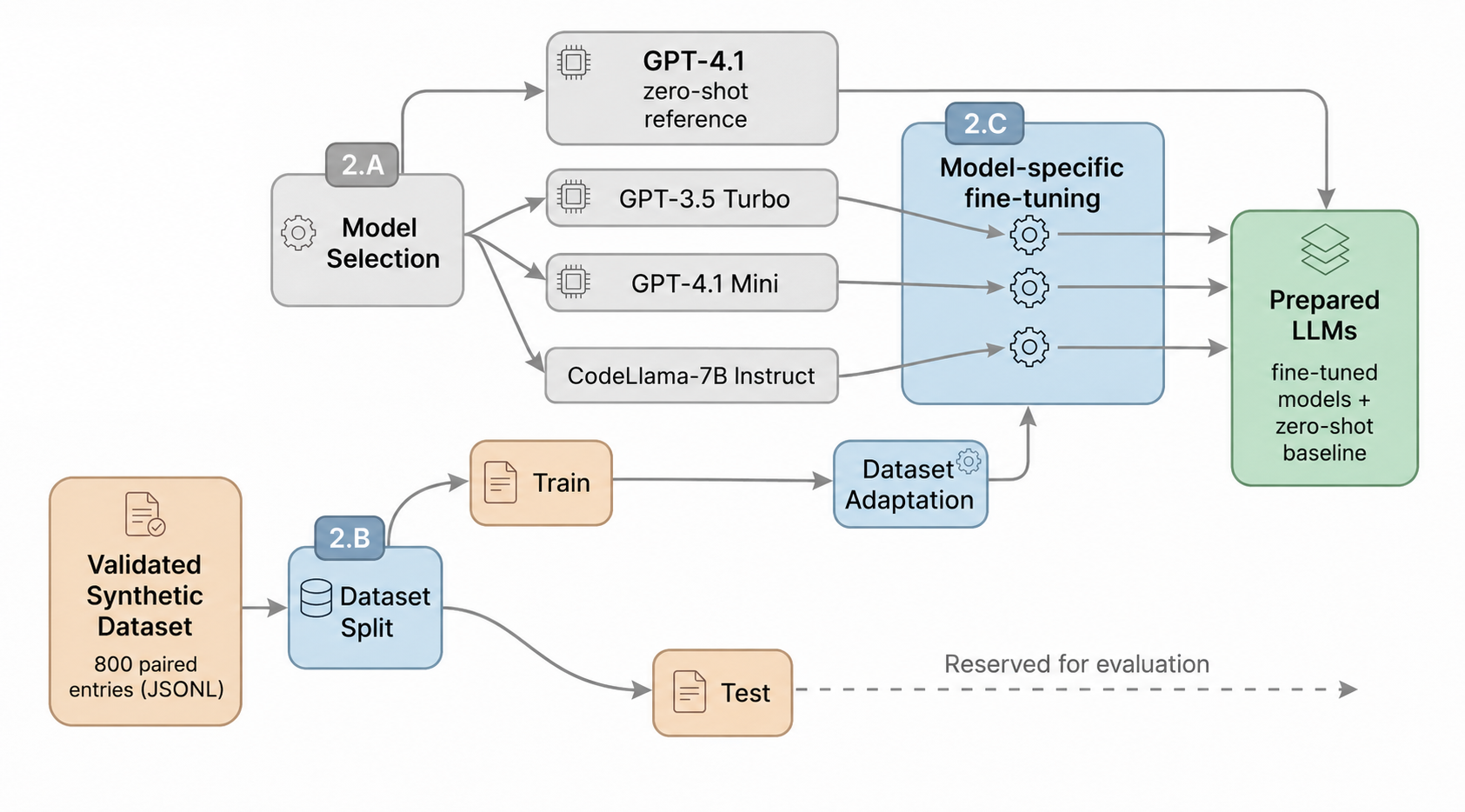}
    \caption{Dynamic results by base case for GPT-4.1-mini (fine-tuned).}
    \label{fig:gpt41mini_dynamic_hist}
\end{figure}

\paragraph{Selection of the best model.} From a qualitative standpoint, the fine-tuned \texttt{GPT-4.1-mini} exhibits a strong ability to preserve structural and stylistic aspects of the original code, maintaining variable names, function signatures and comments intact while transforming only the cryptographic logic. This selectivity is critical for production settings, where unintended edits to business logic could introduce regressions. The model’s capacity to isolate and rewrite only the relevant cryptographic sections demonstrates that LLMs can achieve targeted transformations with high precision once properly specialized.

The comparative results also reveal that model scale is less important than task adaptation. Despite having fewer parameters than the full \texttt{GPT-4.1}, the fine-tuned \texttt{GPT-4.1-mini} outperforms it in all functional metrics (Table~\ref{tab:comparacion-global}). This finding supports the hypothesis that domain-specific fine-tuning yields greater returns than simply increasing model size, particularly when the training data encapsulates consistent patterns of code-to-code transformation.

Overall, the results confirm that cryptographic migration can be automated effectively through instruction-tuned LLMs. Fine-tuned models outperform zero-shot configurations by a wide margin, achieving near-complete correctness in controlled validation. The next section extends this analysis to real-world codebases, testing whether these findings hold under practical conditions involving complex dependencies and unstructured software.

As an additional post-hoc check, we also evaluated a more recent zero-shot model, GPT-5.5, using the same validation split and automated evaluation harness. This experiment was not part of the original model-selection protocol, and is therefore reported only as an exploratory observation. Although GPT-5.5 improved over the GPT-4.1 zero-shot baseline, its results remained far below those obtained by the fine-tuned models, especially GPT-4.1-mini. This observation reinforces the main conclusion of the study: for specialized cryptographic migration tasks, newer general-purpose models may improve zero-shot performance, but domain-specific oriented models remains necessary to achieve reliable functional correctness.

\paragraph{Insights from error analysis.}
Error analysis provides additional insight into the residual failure modes of the fine-tuned \texttt{GPT-4.1-mini}. Fig.~\ref{fig:gpt41_static_hist} breaks down similarity ranges by predominant cryptographic category, making it possible to inspect which families tend to be harder to migrate precisely. Likewise, Fig.~\ref{fig:gpt41mini_dynamic_hist} presents dynamic results by base case, highlighting that the remaining functional failures are concentrated mainly in structurally complex cases, especially key exchange and combined primitive migrations.

Manual inspection of the failing validation cases shows that most residual errors are local software-engineering issues rather than failures to identify the intended cryptographic migration. For instance, variation~6.52, a combined symmetric-cipher and digital-signature case, failed during the post-quantum phase because the generated code referenced signing key material through local variables such as \texttt{priv} and \texttt{pub}, without exposing the namespace bindings expected by the validation harness. In variation~6.57, a combined symmetric-cipher and key-exchange case, the generated code misused the KEM API by passing an invalid public-key buffer to \texttt{oqs.encap\_secret}, producing a \texttt{ValueError: byte string too long}. Additional failures included function-signature mismatches in key-exchange code and assertion-level regressions in MAC self-tests.

These examples, illustrated in ~\ref{app:gpt41mini-errors}, suggest that the main residual weaknesses of the fine-tuned model involve API contracts, namespace management, function signatures, and key-material representation. Importantly, such errors are typically localized and detectable through lightweight validation rules. This suggests that rule-based sanity checks or post-generation repair steps could further increase robustness without changing the underlying fine-tuned model.

\section{Application to real-world cases}
\label{sec:real-world}

To evaluate external validity and practical feasibility, the fine-tuned \texttt{GPT-4.1-mini} model was applied to six real-world open-source repositories containing non-trivial cryptographic logic. Unlike the synthetic dataset, these projects include heterogeneous code styles, undocumented legacy constructs, and dependencies typical of production environments. The objective was threefold: (i) determine whether the model can identify vulnerable or deprecated primitives, (ii) generate correct post-quantum replacements, and (iii) preserve functional behaviour without requiring task-specific prompts or manual guidance.

\begin{figure}
    \centering
    \includegraphics[width=\linewidth]{ 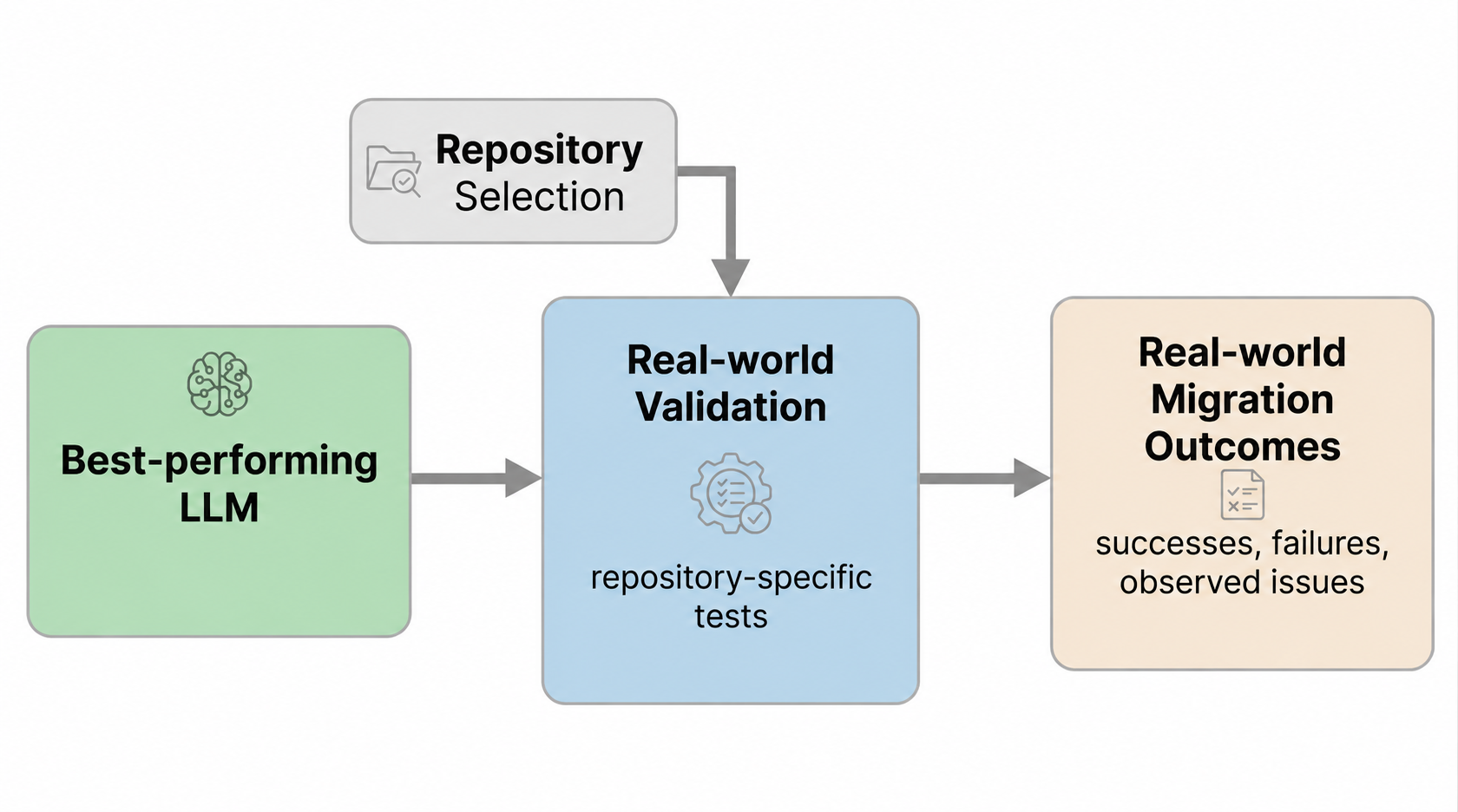}
    \caption{\textit{Real-world Validation Phase}. The best-performing LLM selected in the previous phase is applied to a set of selected open-source repositories. Repository-specific tests are then used to assess practical migration behavior under realistic dependency and modularity constraints, producing a structured summary of real-world migration outcomes.}
    \label{fig:realworld_phase}
\end{figure}

This stage completes the full evaluation pipeline, transitioning from controlled synthetic testing to practical assessment in operational codebases. The workflow for this phase is summarised in Fig.~\ref{fig:realworld_phase}, which illustrates how the top-performing model advances into real-world validation.

\paragraph{Real-world repository selection.}
\label{sec:realworld-selection}

Beyond controlled evaluation on a synthetic corpus, a complementary validation phase was designed to assess the behaviour of the proposed system under realistic software-engineering conditions. The motivation for this stage is twofold. First, real-world codebases exhibit characteristics that are absent from curated datasets, such as heterogeneous coding styles, undocumented assumptions, implicit dependencies, and legacy abstractions. Second, practical post-quantum migration efforts in industry will inevitably operate over such unstructured environments rather than over idealised examples.

The repositories selected for this phase were therefore chosen to strike a balance between realism and experimental tractability. Only open-source projects with demonstrable relevance and sustained community adoption were considered, ensuring that the evaluated code reflects patterns commonly found in production systems. At the same time, the scope was restricted to projects with a manageable code footprint, so that individual cryptographic files could be processed within the effective context window of current large language models without requiring extensive manual decomposition.

A further selection criterion was the explicit use of cryptographic primitives covered during the construction of the synthetic dataset. Concretely, all selected repositories rely on primitives implemented through the \texttt{pyca/cryptography} library and include at least one of the following categories: digital signatures, key exchange, symmetric encryption, hashing, message authentication codes (MAC), or authenticated encryption. This alignment ensures that the real-world evaluation tests generalisation rather than extrapolation beyond the model’s training domain, allowing performance differences to be attributed to structural complexity rather than to unseen cryptographic concepts.

For each repository, a specific source file containing pre-quantum cryptographic logic was identified and used as the direct input to the fine-tuned \texttt{GPT-4.1-mini} model. Unlike the synthetic dataset, these files are not pre-structured, normalised, or automatically categorised. As a result, the cryptographic category relevant to each fragment was manually identified through code inspection and primitive analysis. While this manual labelling step limits scalability, it guarantees correctness during evaluation and mirrors the current state of enterprise migration workflows, where cryptographic inventory is often performed explicitly before remediation.

The evaluation of real-world migrations focuses on three complementary dimensions. First, local precision assesses whether the model modifies only the cryptographic components that require migration, preserving surrounding business logic and control flow. Second, cryptographic correctness evaluates whether deprecated or quantum-vulnerable primitives are replaced by appropriate post-quantum or quantum-resistant alternatives consistent with current guidance. Finally, functional compatibility measures whether the migrated code remains syntactically valid, can be loaded by the Python interpreter, and executes correctly under repository-specific smoke tests, understood here as lightweight functional checks that exercise the relevant cryptographic execution paths without requiring the full original test suite. These tests allow the evaluation to detect runtime regressions while keeping the real-world validation phase tractable.

Taken together, these selection and evaluation criteria ensure that the repositories provide a meaningful and demanding test bed for LLM-assisted cryptographic migration. The final selection, summarized in Table~\ref{tab:real-cases}, includes open-source projects with active adoption, manageable cryptographic files, use of primitives represented in the synthetic dataset, and relevance to realistic software-engineering scenarios.

\begin{table*}[t]
\centering
\caption{Real-world repositories used for external validation}
\label{tab:real-cases}
\renewcommand{\arraystretch}{1.15}
\begin{tabular}{|p{2.5cm}|p{2.7cm}|p{1.5cm}|p{5cm}|}
\hline
\textbf{Project} & \textbf{Category} & \textbf{LOC} & \textbf{Primitives identified} \\ \hline
Netflix Lemur & Certificate management & 130 & RSA, ECDSA, SHA-1/256 \\ \hline
python-jose & JWT signing and encryption & 586 & RSA, ECDSA, HMAC, AES-GCM \\ \hline
Prefect & Secret encryption in workflows & 49 & AES-128-CBC, HMAC-SHA256 \\ \hline
SnapPass & One-time password sharing & 362 & Fernet (AES-CBC + HMAC) \\ \hline
google-auth & Token signing and verification & 151 & RSA-PSS, SHA-256/512 \\ \hline
Alexa SDK & HTTP signature verification & 511 & RSA-SHA256, X.509 \\ \hline
\end{tabular}
\end{table*}

\paragraph{Methodology applied.}
The process follows a uniform and reproducible pipeline to apply the trained model to real repositories and to validate the resulting migrations under realistic constraints. Concretely:

\begin{enumerate}
    \item \textbf{Load the original source file:} for each repository, the specific file containing cryptographic logic to be migrated was identified and loaded from its original location in the public repository.

    \item \textbf{Split into fragments (tokenisation):} since \texttt{GPT-4.1-mini} supports an effective context window of $\sim$8000 tokens (accounting for prompt overhead), each file was automatically split into manageable fragments of approximately 7800 usable tokens using the \texttt{tiktoken} encoder.

    \item \textbf{Prompt construction with explicit category:} for every fragment, a concise prompt was generated following the same structure used during training/validation, explicitly stating the cryptographic category relevant to that fragment (e.g., signatures, AEAD, MAC, hashing). Category assignment was performed manually via code inspection and primitive identification, ensuring accurate labelling but limiting scalability. As future work, this step could be automated using syntactic/static analysis or a dedicated classifier so the full pipeline can operate autonomously on large codebases.

    \item \textbf{Automatic migration via the fine-tuned model:} each prompted fragment was sent to the fine-tuned \texttt{GPT-4.1-mini} model to produce a post-quantum-oriented refactoring. Model outputs were stored as-is (without additional commentary) to facilitate deterministic reconstruction.

    \item \textbf{Reconstruction of migrated code:} migrated fragments were concatenated and reassembled into a single executable module representing the post-quantum version of the original file.

    \item \textbf{Manual and automatic evaluation:} the resulting code was reviewed via smoke tests, automated \texttt{diff} against the original (to surface structural regressions), and a targeted manual inspection to identify and document type-level issues, incorrect encodings, or runtime incompatibilities introduced by the migration.
\end{enumerate}

This methodology standardises real-world evaluation, making outcomes comparable across repositories and enabling systematic identification of recurrent failure modes.

\paragraph{Per-repository results and observed failure modes.} Across all projects, the model detected occurrences of pre-quantum cryptography and proposed PQC-compatible refactorings that generally matched the structure and intent of the original code. The following repository-level results highlight both strengths and limits:

\textbf{Netflix Lemur.}
The model correctly replaced RSA and ECDSA constructs, modernising certificate-handling functions. A recurrent error involved invalid key lengths when switching to \texttt{ChaCha20Poly1305}, which was resolved manually. Aside from this, the migrated file executed successfully, demonstrating high accuracy on PKI-related transformations.

\textbf{python-jose.}
As one of the most complex projects, \texttt{python-jose} exposed the model’s limits. Multiple primitives and interdependent APIs caused structural inconsistencies in the output, leading to non-executable code. The migration produced valid local substitutions but failed to preserve global consistency, indicating that multi-layered dependencies remain challenging for current LLMs.

\textbf{Prefect.}
This repository provided a small, well-contained symmetric module (CBC + HMAC). The model substituted AES-CBC and HMAC-SHA256 for \texttt{ChaCha20Poly1305}, but incorrectly assumed hexadecimal encoding for environment keys. After a minimal manual correction, the program executed normally, suggesting that fine-tuned LLMs can handle compact cryptographic components.

\textbf{SnapPass.}
This project, relying on Fernet, provided a clean AEAD-like scenario. The model produced a syntactically and functionally valid migration, with only one minor error---attempting to decode a binary key as UTF-8. Correcting this line yielded a fully functional post-quantum version. This case confirmed that the system can produce production-ready migrations in simple contexts.

\textbf{google-auth.}
The migration preserved the overall structure but failed during runtime due to incorrect key serialisation. The model occasionally represented keys as integers rather than byte objects, causing exceptions in signature generation. These issues highlight the importance of explicit type handling in future training data.

\textbf{Alexa SDK.}
Here, the only cryptographic operation was hashing for signature verification. The model replaced instances of \texttt{SHA256} with \texttt{SHA3\_512} while maintaining documentation and logic unchanged. The resulting file was syntactically correct and passed all smoke tests without modification, exemplifying a perfect migration.

Overall, these experiments revealed consistent trends: the model performed best in localised, self-contained modules (e.g., SnapPass, Prefect, Alexa SDK) and required minimal human supervision to achieve functional results. Larger frameworks with intertwined cryptographic layers (e.g., \texttt{python-jose}) exposed limitations in global reasoning and dependency tracking.

\begin{table*}[t]
\centering
\caption{Summary of real-world migration outcomes}
\label{tab:real-summary}
\renewcommand{\arraystretch}{1.15}
\begin{tabular}{|p{2.5cm}|p{2.8cm}|p{2cm}|p{6cm}|}
\hline
\textbf{Project} & \textbf{Main category} & \textbf{Result} & \textbf{Issues observed} \\ \hline
Lemur & PKI / certificate management & Partial success & Incorrect key size; missing exception handling \\ \hline
python-jose & Mixed (sign, MAC, AEAD) & Failure & Structural inconsistency; dependency conflicts \\ \hline
Prefect & Authenticated encryption & Partial success & Misinterpreted key encoding \\ \hline
SnapPass & Authenticated encryption & Full success & Minor decoding error \\ \hline
google-auth & Digital signatures & Failure & Type mismatch for key objects \\ \hline
Alexa SDK & Hashing & Full success & None detected \\ \hline
\end{tabular}
\end{table*}

\paragraph{Execution-level validation and operational considerations.}
Beyond correctness, the real-world evaluation assessed operational efficiency and the practical role of the model within a realistic migration workflow. Unlike synthetic benchmarks, production repositories require validation strategies robust to incomplete documentation, external dependencies, and legacy abstractions.

Evaluation was performed by isolating representative execution paths for each project. For every repository, a small set of application-level functions exercising the cryptographic logic (e.g., token generation and verification, encryption--decryption cycles, signature validation) was identified and executed on the original pre-quantum code. After migration, the same functions were re-executed on the transformed code and their outputs compared. Any deviation in behaviour, exception handling, or output structure was flagged as a functional regression. This approach evaluates correctness independently of surrounding application logic while remaining faithful to real usage patterns.

From an efficiency standpoint, the migration required no repository-level prompt engineering or task-specific tuning. Each fragment was processed using the same inference configuration and prompting strategy validated during synthetic testing, confirming that the fine-tuned model generalises without additional overhead. The average inference time per fragment remained within the same order of magnitude as the validation phase, and no post-processing beyond lightweight key/type fixes (as described above) was required. This suggests the approach can scale to real codebases without prohibitive computational or operational cost.

\paragraph{Integrating LLM-assisted code migration into cryptographic transition frameworks.} 
A key outcome of this work is the confirmation that LLM-assisted code migration aligns closely with enterprise-level cryptographic transition frameworks, particularly in bridging the gap between policy-level decisions and code-level remediation.

Enterprise initiatives such as ELCA \cite{sikeridis2023elca} and the framework proposed by Hasan et al. \cite{hasan2024framework} emphasize the need for structured transition pipelines involving inventory, dependency mapping, algorithm selection, and eventual migration. Likewise, industry-led programmes—including IBM’s Cryptography Bill of Materials (CBOM) \cite{ibm2024cbom} and CISA’s Automated Cryptographic Discovery and Inventory (ACDI) \cite{cisa2024pqcstrategy}—focus on detecting and classifying vulnerable cryptographic components in codebases. However, these efforts stop short of automated code transformation.

This project contributes directly at the “Remediation” phase of such pipelines. After discovery tools identify what needs to change, our LLM-based system answers how to change it. In IBM’s CBOM workflow, this corresponds to the final “Remediate” stage, which involves adapting software to conform to updated cryptographic standards—something that today often relies on manual rewriting or ad hoc scripting. The experiments presented here show that a fine-tuned model can act as an automated cryptographic refactoring agent, capable of detecting deprecated primitives (e.g., RSA, ECDSA, AES-128), proposing compliant post-quantum replacements (e.g., Dilithium, Kyber, AES-256), and preserving functional behaviour across real-world projects.

\begin{figure}
    \centering
    \includegraphics[width=1\linewidth]{ 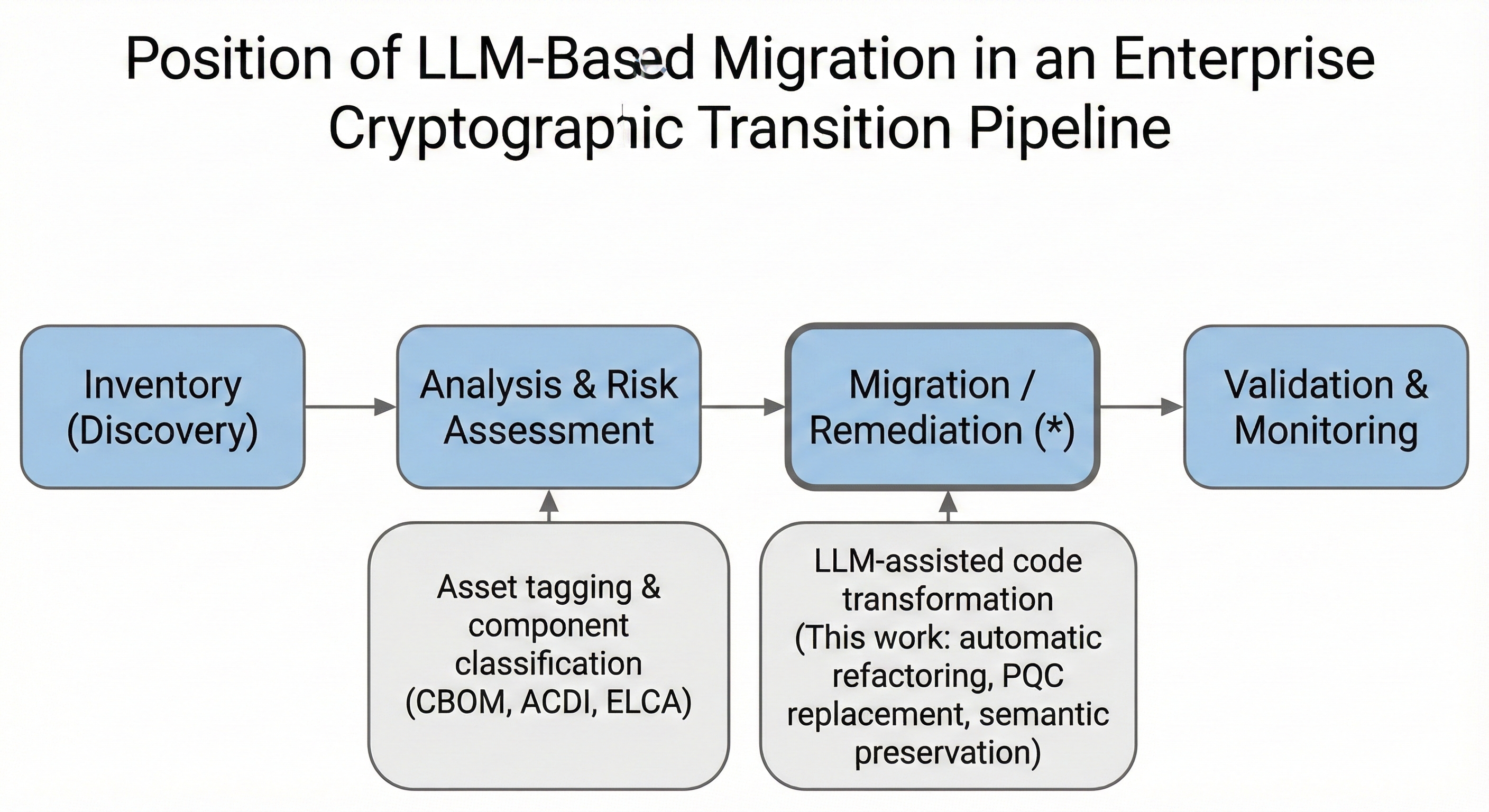}
    \caption{Position of the proposed LLM-assisted migration system within a standardized enterprise cryptographic transition pipeline.}
    \label{fig:framework}
\end{figure}

This practical capability addresses the needs articulated by ENISA \cite{ENISA2021PQC} and NIST \cite{nist_cswp_39_2025} for crypto-agile systems, defined as systems that can evolve cryptographically without overhauling the entire software stack. Rather than merely reporting issues, the system operationalises crypto-agility by carrying out the code modifications required by policy. In doing so, it fills a long-standing gap between governance and engineering. This alignment supports contribution C4, showing that fine-tuned LLMs can serve as concrete migration engines within broader organizational PQC strategies, making cryptographic modernization more scalable, less error-prone, and accessible to security engineering teams. 
Fig.~\ref{fig:framework} illustrates the position of this work within a canonical enterprise cryptographic migration workflow.

\section{Discussion and limitations}
\label{sec:discussion-limitations}

This work is intentionally scoped to enable verifiable, reproducible experimentation within realistic constraints. While this enables meaningful results in current software environments, it also imposes clear boundaries on generalisability. The most relevant design choices and their implications are discussed below.

\begin{enumerate}[label=\textbf{\arabic*.}, leftmargin=*]

\item \textbf{Language and library scope.}
The decision to focus on Python and the \texttt{PyCA/Cryptography} library ensures a stable and widely supported substrate. Python’s dominance in academic and production settings enhances reproducibility, while \texttt{PyCA} offers a well-structured, security-focused API that minimizes ambiguity in usage. These factors enable consistent functional testing and streamline the migration process.

\item \textbf{Primitive-level coverage.}
The cryptographic primitives targeted are those explicitly affected by quantum threats. Classical asymmetric mechanisms (e.g., RSA, ECDSA, DH) are migrated to post-quantum counterparts (e.g., Dilithium, Kyber), while symmetric and hash functions are updated via parameter strengthening (e.g., AES key size increases, SHA-2 to SHA-3). This aligns with NIST and ENISA guidance for hybrid transitions. It is also worth exploring migration to other post-quantum primitives that have been standardized in a second batch -- such as HQC --- or are currently under exploration -- as it is the case with the signature schemes being evaluated towards standardization by NIST, see~\cite{NIST_sig}.

\item \textbf{Emphasis on authenticated encryption.}
Although historically overlooked in foundational taxonomies, AEAD constructions are critical in modern protocols such as TLS~1.3 and QUIC. This work treats AEAD as a first-class component, preserving its usage while replacing the key-exchange layer with KEM constructions. This mirrors real-world post-quantum deployment strategies, including NIST’s TLS migration trials.

\item \textbf{Protocol and abstraction boundaries.}
The system does not aim to migrate full communication protocols (e.g., TLS, SSH) or higher-level constructs like handshake flows or network abstractions. 
It also excludes cryptographic constructions that are closely related to the primary schemes under analysis—such as threshold signatures or key exchange protocols derived from KEMs—as well as more advanced cryptographic primitives (e.g., zero-knowledge proofs, multiparty computation protocols, and commitment schemes), whose correctness remains beyond the current capabilities of LLMs.

\item \textbf{Library and ecosystem dependency.}
The migrations rely on existing PQC implementations such as \texttt{liboqs-python}. While this reflects current industry best practices, it means that the approach inherits the constraints and assumptions of these tools. As standards evolve and new implementations are adopted, the models may require retraining to reflect updated APIs and security profiles.

\item \textbf{Contextual and cross-module limitations.}
The system performs reliably on isolated or well-encapsulated cryptographic functions. However, in real-world scenarios involving distributed logic or interdependent abstractions, local transformations can break global consistency. These errors are difficult to detect via unit tests alone and suggest a need for integrating dependency analysis or symbolic reasoning in future work.

\end{enumerate}

In sum, the design constraints are not merely limiting—they are enabling. They allow the problem of post-quantum migration to be framed in a way that supports automatic transformation and functional verification with current LLM technology. While generalising to broader cryptographic domains remains open, this work provides a concrete and reproducible foundation upon which such efforts can build.

\section{Conclusions and future work}
\label{sec:conclusions}

This study investigated whether large language models can assist in the migration of cryptographic code from pre-quantum to post-quantum standards in a verifiable and efficient manner. The initial research question posed in this work was:

\begin{quote}
\emph{Can large language models be effectively trained and evaluated to assist in the migration of pre-quantum cryptographic code to post-quantum counterparts while preserving functional correctness?}
\end{quote}

\begin{figure}
    \centering
    \includegraphics[width=1\linewidth]{ 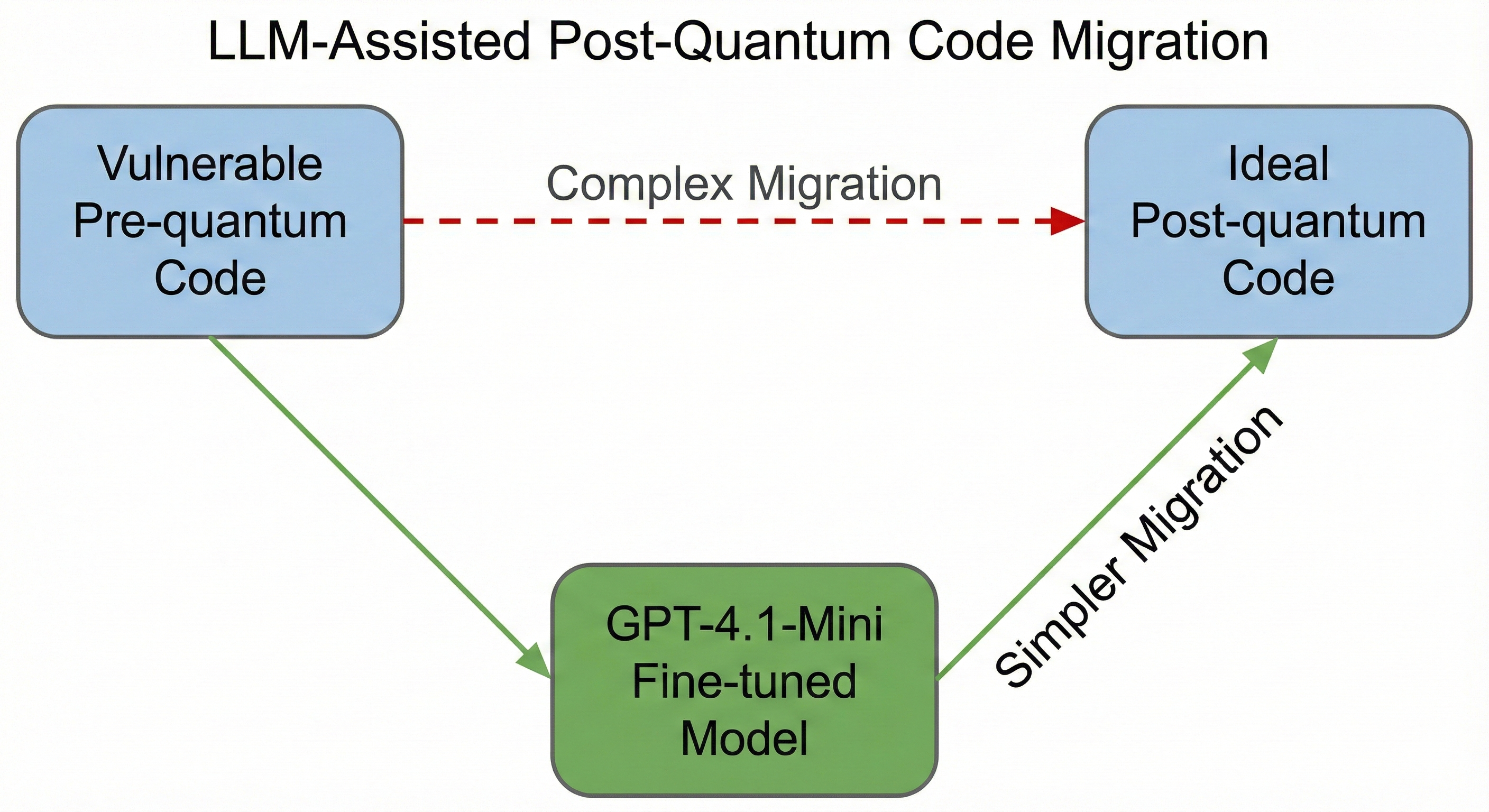}
    \caption{GPT-4.1-Mini–assisted post-quantum migration, fine-tuned, reduces process complexity and serves as an effective alternative to the manual approach.}
    \label{fig:conclusion}
\end{figure}

The results of this work provide a clear and affirmative answer.  
Through a combination of a carefully curated dataset of 800 paired examples and a rigorous fine-tuning and validation pipeline, the evidence shows that LLMs can indeed perform functionally correct and structurally consistent cryptographic migrations across a wide variety of primitives. As illustrated in Fig.~\ref{fig:conclusion}, the proposed methodology substantially streamlines the migration workflow, reducing manual intervention and transforming what was previously a multi-stage, error-prone process into a guided, verifiable pipeline powered by a fine-tuned LLM.

The proposed system integrated two core components: (i) a structured dataset covering six major cryptographic families and combined primitive cases, and (ii) a fine-tuned model optimized for cryptographic transformations while preserving functional semantics. This combination enabled controlled experimentation across synthetic benchmarks and real open-source repositories, offering a comprehensive view of LLM-assisted modernization.

From the experimental results, several conclusions emerge.

First, \textbf{domain-specific fine-tuning is essential}. Zero-shot models, despite their fluency, failed to achieve reliable cryptographic correctness. After specialization, the fine-tuned \texttt{GPT-4.1-mini} achieved an average similarity of 0.91 and passed 92.5\% of dynamic tests, outperforming larger general-purpose models and the CodeLlama-7B baseline.

Second, the system shows \textbf{remarkable cost-efficiency}: the best-performing GPT-4.1-mini configuration required less than \$6 in API usage for training and evaluation, making this approach accessible for academic research and enterprise environments.

Third, the fine-tuned model preserved \textbf{structural fidelity}: edits remained confined to cryptographic sections, and surrounding logic, comments and control flow remained intact. This selective rewriting is critical for real-world engineering use cases, where unintended modifications may introduce regressions or security flaws.

Applying the model to real-world repositories further validated its external applicability. In four of the six evaluated projects—such as \emph{SnapPass}, \emph{Prefect}, and the \emph{Alexa SDK}—the migrated code was functional or nearly functional with minimal manual intervention. More complex projects, including \emph{google-auth} and \emph{python-jose}, exposed current limitations, especially when dealing with intertwined abstractions, implicit dependencies or multi-layered cryptographic flows. These observations highlight that LLMs excel at local migrations but still struggle with architecture-wide reasoning, reinforcing the need for symbolic validation and dependency-aware pipelines.

Taken together, the empirical findings demonstrate that the answer to the research question is \textbf{yes}: large language models \emph{can} be effectively trained and evaluated to support the migration of pre-quantum cryptographic code to post-quantum counterparts while maintaining functional correctness. This is feasible at low cost, with reproducible validation, and with a level of precision that makes LLM-assisted cryptographic modernization a realistic tool for engineering teams. The experiments confirm that such models do not merely rewrite code superficially—they perform semantically meaningful transformations that pass runtime verification.

Looking forward, several research directions emerge:
(i) \textbf{Automatic fragment classification}. Integrating a classifier to detect cryptographic categories would eliminate manual tagging and enable autonomous full-repository scanning.
(ii) \textbf{Dataset expansion}. Incorporating curated examples from industrial systems would improve robustness against complex dependency graphs and multi-file migrations.
(iii) \textbf{Post-processing enhancements}. Embedding sanity checks for key sizes, imports, encodings and nonce management would further reduce structural errors.
(iv) \textbf{Multi-agent orchestration}. Coordinating different LLMs for generation, verification and documentation could improve reliability and explainability.
(v) \textbf{Automated dynamic testing}. Generating and running validation scripts automatically from model outputs would create a closed-loop verification cycle.
(vi) \textbf{Language diversification}. Extending the methodology to languages such as Go, C or Java would test its scalability beyond Python and enable broader adoption.
(vii) \textbf{Migration of protected data at rest}. While this work focuses on source-code migration, a complete organizational PQC transition must also address data already encrypted or authenticated with legacy primitives. Future work should therefore explore how code-level migration can be coordinated with data re-encryption, re-authentication, key rotation, and backward-compatible access strategies for long-lived stored data.

In summary, this work establishes a reproducible, data-driven framework for post-quantum cryptographic migration using large language models. A compact, fine-tuned model such as \texttt{GPT-4.1-mini} can perform accurate, cost-effective migrations at scale. This confirms that LLMs can serve as practical assistants for transitioning legacy infrastructures toward quantum-safe standards, marking a tangible step toward secure and future-proof cryptographic ecosystems. The findings lay the groundwork for hybrid pipelines where AI collaborates with engineers to produce reliable, verifiable and crypto-agile software systems.

\section*{Declarations}

\subsection*{Funding}
This work was supported by a Collaboration Grant in University Departments for the 2025--2026 academic year, awarded by the Spanish Ministry of Education, Vocational Training and Sports. The work of A.I.G.-T. is supported by projects PID2022-140126OB-I00 and PID2023-150310OB-I00, and the work of M.I.G.V. by project PID2023-151238OA-I00, all the projects funded by MCIU/AEI/FEDER UE, Spain. In addition, the work of J.P.B. and A.I.G.-T. was supported by the EMACS cybersecurity research network (Project RED2024-123456-T, funded by MICIU/AEI/10.13039/501100011033).

\subsection*{Competing interests}
The authors declare that they have no competing interests.

\subsection*{Ethics approval}
Not applicable. This study did not involve human participants, human data, or animals.

\subsection*{Consent to participate}
Not applicable.

\subsection*{Consent for publication}
Not applicable.

\subsection*{Data availability}
The dataset generated and analysed during the current study is publicly available in the e-cienciaDatos repository under the title \textit{Cryptographic Migration Dataset: Pre-Quantum to Post-Quantum}, version V1, with DOI \href{https://doi.org/10.21950/7GK4MJ}{10.21950/7GK4MJ}~\cite{7GK4MJ_2025}.

\subsection*{Code availability}
The code used to generate, fine-tune, evaluate, and analyse the models is not publicly available at this stage due to ongoing development and repository cleanup. The core scripts required to reproduce the evaluation results can be made available from the corresponding author upon reasonable request. A public release of the evaluation harness and reproducibility scripts is planned for a future version of the artifact.

\subsection*{Author contributions}
Javier Pallarés de Bonrostro led the technical development of the work, including dataset construction, software implementation, model fine-tuning, validation, experimental analysis, visualization, and writing of the original draft. 
Ana Isabel González-Tablas originated the research idea, contributed to the conceptualization and methodological design of the study, supervised the work on a day-to-day basis, contributed to funding acquisition and project administration, and participated in manuscript review and editing. 
María Isabel González Vasco contributed expert knowledge in post-quantum cryptography, provided relevant conceptual and methodological input, supported the cryptographic analysis and validation of the migration rationale, contributed to funding acquisition, and participated in manuscript review and editing.

\subsection*{Use of generative AI and AI-assisted technologies}
Generative AI tools were used as part of the research object of this study, namely for the generation and evaluation of cryptographic migration examples as described in the methodology. All scientific claims, code, experimental results, citations, and final text were reviewed and validated by the authors, who remain fully responsible for the content of the manuscript.

\printbibliography

\onecolumn

\appendix

\section{Full dataset distribution}
\label{app:dataset}

\begin{table*}[h]
\centering
\caption{Distribution of single cryptographic primitives across training and validation splits.}
\label{tab:dataset_split_primitives} 
\begin{minipage}{0.48\textwidth}
\centering
\begin{tabular}{p{3.2cm} r r r}
\toprule
\textbf{Primitive / Alg.} & \textbf{\#Var.} & \textbf{Train} & \textbf{Val.} \\
\midrule

\multicolumn{4}{l}{\textbf{Hash functions}}\\
\addlinespace[2pt]
BLAKE2b        & 7  & 7  & 0 \\
BLAKE2s        & 7  & 7  & 0 \\
MD5            & 10 & 9  & 1 \\
RIPEMD160      & 1  & 1  & 0 \\
SHA-1          & 14 & 13 & 1 \\
SHA-224        & 11 & 10 & 1 \\
SHA-256        & 14 & 12 & 2 \\
SHA-384        & 12 & 12 & 0 \\
SHA-512        & 12 & 11 & 1 \\
SHA-512/224    & 6  & 4  & 2 \\
SHA-512/256    & 6  & 4  & 2 \\
\addlinespace[6pt]
\multicolumn{4}{l}{\textbf{Symmetric encryption}}\\
\addlinespace[2pt]
3DES           & 7  & 6  & 1 \\
AES-128        & 82 & 74 & 8 \\
DES            & 11 & 10 & 1 \\
\addlinespace[6pt]

\multicolumn{4}{l}{\textbf{Message authentication codes (MAC)}}\\
\addlinespace[2pt]
CMAC-AES-128               & 12 & 11 & 1 \\
CMAC-AES-256               & 4  & 3  & 1 \\
CMAC-3DES                  & 2  & 2  & 0 \\
GMAC-AES-128               & 5  & 5  & 0 \\
GMAC-AES-256               & 3  & 3  & 0 \\
HKDF-SHA-256 + Poly1305     & 1  & 1  & 0 \\
HKDF-SHA-256 + HMAC-SHA-256 & 1  & 1  & 0 \\
HMAC-BLAKE2b               & 2  & 2  & 0 \\
HMAC-BLAKE2b-256           & 1  & 1  & 0 \\
HMAC-BLAKE2s               & 3  & 3  & 0 \\
HMAC-MD5                   & 1  & 1  & 0 \\
HMAC-SHA-1                 & 11 & 9  & 2 \\
HMAC-SHA-224               & 4  & 4  & 0 \\
HMAC-SHA-256               & 37 & 32 & 5 \\
HMAC-SHA-512               & 1  & 1  & 0 \\
HMAC-SM3                   & 2  & 2  & 0 \\
PBKDF2 + HMAC-SHA-1        & 2  & 1  & 1 \\
Poly1305                   & 7  & 7  & 0 \\
\addlinespace[6pt]

\bottomrule
\end{tabular}

\subcaption{First part}
\end{minipage}
\hfill
\begin{minipage}{0.48\textwidth}
\centering

\begin{tabular}{p{3.2cm} r r r}
\toprule
\textbf{Primitive / Alg.} & \textbf{\#Var.} & \textbf{Train} & \textbf{Val.} \\
\midrule

\multicolumn{4}{l}{\textbf{Authenticated encryption (AEAD)}}\\
\addlinespace[2pt]
AES-128-CCM     & 24 & 21 & 3 \\
AES-128-GCM     & 37 & 35 & 2 \\
AES-192-GCM     & 4  & 4  & 0 \\
AES-192-CCM     & 1  & 0  & 1 \\
AES-256-CCM     & 2  & 2  & 0 \\
AES-256-GCM     & 1  & 0  & 1 \\
AES-256-GCM-SIV & 7  & 6  & 1 \\
AES-256-SIV     & 15 & 14 & 1 \\
AES-512-SIV     & 1  & 1  & 0 \\
AES-OCB3        & 6  & 6  & 0 \\
Fernet          & 2  & 1  & 1 \\
\addlinespace[6pt]

\multicolumn{4}{l}{\textbf{Digital signatures}}\\
\addlinespace[2pt]
DSA             & 17 & 14 & 3 \\
ECDSA-SECP256R1  & 24 & 22 & 2 \\
ECDSA-SECP384R1  & 4  & 3  & 1 \\
Ed25519          & 18 & 17 & 1 \\
Ed448            & 11 & 9  & 2 \\
RSA              & 26 & 25 & 1 \\
\addlinespace[6pt]

\multicolumn{4}{l}{\textbf{Key exchange}}\\
\addlinespace[2pt]
DH-2048         & 9  & 9  & 0 \\
DH-3072         & 7  & 7  & 0 \\
ECDH-P384       & 4  & 4  & 0 \\
ECDH-SECP256R1  & 22 & 20 & 2 \\
ECDH-SECP384R1  & 7  & 5  & 2 \\
ECDH-SECP521R1  & 4  & 3  & 1 \\
ECIES           & 8  & 7  & 1 \\
RSA-2048        & 9  & 8  & 1 \\
X25519          & 24 & 22 & 2 \\
X448            & 6  & 5  & 1 \\

\bottomrule
\end{tabular}
\subcaption{Second part}
\end{minipage}
\end{table*}



\begin{table*}[h]
\centering
\caption{Distribution of combined constructions across training and validation splits.}
\label{tab:dataset_split_composites}
\begin{minipage}{0.48\textwidth}
\centering
\begin{tabular}{p{3.2cm} r r r}

\toprule
\textbf{Construction} & \textbf{\#Var.} & \textbf{Train} & \textbf{Val.} \\
\midrule

\multicolumn{4}{l}{\textbf{AEAD + digital signature}}\\
\addlinespace[2pt]
AES-GCM + ECDSA-P-256              & 4 & 4 & 0 \\
AES-GCM + RSA-2048                 & 4 & 4 & 0 \\
ChaCha20-Poly1305 + ECDSA-P-256    & 4 & 4 & 0 \\
ChaCha20-Poly1305 + RSA-2048       & 4 & 3 & 1 \\
\addlinespace[6pt]

\multicolumn{4}{l}{\textbf{AEAD + hashing}}\\
\addlinespace[2pt]
AES-GCM + SHA-256                  & 4 & 4 & 0 \\
AES-GCM + SHA-512                  & 4 & 4 & 0 \\
ChaCha20-Poly1305 + SHA-256        & 4 & 4 & 0 \\
ChaCha20-Poly1305 + SHA-512        & 4 & 3 & 1 \\
\addlinespace[6pt]

\multicolumn{4}{l}{\textbf{AEAD + key exchange}}\\
\addlinespace[2pt]
AES-GCM + Diffie-Hellman-3072      & 4 & 4 & 0 \\
AES-GCM + X25519                   & 3 & 3 & 0 \\
ChaCha20-Poly1305 + Diffie-Hellman-3072 & 4 & 3 & 1 \\
ChaCha20-Poly1305 + X25519         & 5 & 4 & 1 \\
\addlinespace[6pt]

\multicolumn{4}{l}{\textbf{AEAD + MAC}}\\
\addlinespace[2pt]
AES-GCM + HMAC-SHA-256             & 4 & 4 & 0 \\
ChaCha20-Poly1305 + HMAC-SHA-256   & 4 & 3 & 1 \\
\addlinespace[6pt]

\multicolumn{4}{l}{\textbf{Hashing + digital signature}}\\
\addlinespace[2pt]
SHA-256 + ECDSA-P-256              & 5 & 4 & 1 \\
SHA-256 + RSA-2048                 & 4 & 4 & 0 \\
SHA-512 + ECDSA-P-256              & 4 & 4 & 0 \\
SHA-512 + RSA-2048                 & 3 & 3 & 0 \\
\addlinespace[6pt]

\multicolumn{4}{l}{\textbf{Hashing + key exchange}}\\
\addlinespace[2pt]
SHA-256 + Diffie-Hellman-3072      & 4 & 4 & 0 \\
SHA-256 + X25519                   & 4 & 3 & 1 \\
SHA-512 + Diffie-Hellman-3072      & 4 & 4 & 0 \\
SHA-512 + X25519                   & 4 & 4 & 0 \\
\addlinespace[6pt]

\multicolumn{4}{l}{\textbf{Hashing + MAC}}\\
\addlinespace[2pt]
SHA-256 + HMAC-SHA-256             & 4 & 3 & 1 \\
SHA-512 + HMAC-SHA-256             & 4 & 4 & 0 \\
\addlinespace[6pt]

\bottomrule
\end{tabular}

\subcaption{First part}
\end{minipage}
\hfill
\begin{minipage}{0.48\textwidth}
\centering

\begin{tabular}{p{3.2cm} r r r}

\toprule
\textbf{Construction} & \textbf{\#Var.} & \textbf{Train} & \textbf{Val.} \\
\midrule

\multicolumn{4}{l}{\textbf{MAC + digital signature}}\\
\addlinespace[2pt]
HMAC-SHA-256 + ECDSA-P-256         & 4 & 4 & 0 \\
HMAC-SHA-256 + RSA-2048            & 4 & 3 & 1 \\
\addlinespace[6pt]

\multicolumn{4}{l}{\textbf{MAC + key exchange}}\\
\addlinespace[2pt]
HMAC-SHA-256 + Diffie-Hellman-3072 & 4 & 3 & 1 \\
HMAC-SHA-256 + X25519              & 4 & 4 & 0 \\
\addlinespace[6pt]

\multicolumn{4}{l}{\textbf{Symmetric cipher + AEAD}}\\
\addlinespace[2pt]
AES-128 + AES-GCM                  & 4 & 3 & 1 \\
AES-128 + ChaCha20-Poly1305        & 4 & 4 & 0 \\
AES-256 + AES-GCM                  & 4 & 4 & 0 \\
AES-256 + ChaCha20-Poly1305        & 4 & 3 & 1 \\
\addlinespace[6pt]

\multicolumn{4}{l}{\textbf{Symmetric cipher + digital signature}}\\
\addlinespace[2pt]
AES-128 + ECDSA-P-256              & 4 & 4 & 0 \\
AES-128 + RSA-2048                 & 4 & 4 & 0 \\
AES-256 + ECDSA-P-256              & 4 & 4 & 0 \\
AES-256 + RSA-2048                 & 4 & 2 & 2 \\
\addlinespace[6pt]

\multicolumn{4}{l}{\textbf{Symmetric cipher + hashing}}\\
\addlinespace[2pt]
AES-128 + SHA-256                  & 4 & 4 & 0 \\
AES-128 + SHA-512                  & 4 & 4 & 0 \\
AES-256 + SHA-256                  & 4 & 4 & 0 \\
AES-256 + SHA-512                  & 4 & 3 & 1 \\
\addlinespace[6pt]

\multicolumn{4}{l}{\textbf{Symmetric cipher + key exchange}}\\
\addlinespace[2pt]
AES-128 + Diffie-Hellman-3072      & 4 & 3 & 1 \\
AES-128 + X25519                   & 4 & 4 & 0 \\
AES-256 + Diffie-Hellman-3072      & 4 & 4 & 0 \\
AES-256 + X25519                   & 4 & 3 & 1 \\
\addlinespace[6pt]

\multicolumn{4}{l}{\textbf{Symmetric cipher + MAC}}\\
\addlinespace[2pt]
AES-128 + HMAC-SHA-256             & 4 & 3 & 1 \\
AES-256 + HMAC-SHA-256             & 4 & 4 & 0 \\

\bottomrule
\end{tabular}
\subcaption{Second part}

\end{minipage}
\end{table*}

\clearpage
\twocolumn

\section{Representative GPT-4.1-mini failure examples}
\label{app:gpt41mini-errors}

Listing~\ref{lst:gpt41mini-error-examples} shows two representative runtime failures observed in the outputs generated by the fine-tuned \texttt{GPT-4.1-mini}. Both examples correspond to raw model outputs evaluated through the dynamic validation harness. The first failure illustrates a namespace-binding problem in a combined symmetric-cipher and digital-signature migration. The second illustrates a KEM API misuse in a combined symmetric-cipher and key-exchange migration.

\begin{lstlisting}[
basicstyle=\ttfamily\scriptsize,
frame=single,
breaklines=true,
caption={Representative residual failures observed in fine-tuned \texttt{GPT-4.1-mini} outputs.},
label={lst:gpt41mini-error-examples}
]
# Variation 6.52: missing namespace bindings
# Category: symmetric cipher + digital signature

with oqs.Signature(ALGO) as s:
    s.secret_key = ctypes.create_string_buffer(priv, len(priv))
    sig = s.sign(ct)

# Harness traceback:
# RuntimeError: handle_cipher_sign_auto:
# missing SYM_KEY/SYM, _PRIV/PRIV, or _PUB/PUB in namespace


# Variation 6.57: KEM API misuse
# Category: symmetric cipher + key exchange

with oqs.KeyEncapsulation(ALGORITHM) as kem:
    kem.secret_key = ct.create_string_buffer(priv, len(priv))
    _, shared = kem.encap_secret(peer_pub)

# Runtime traceback:
# ValueError: byte string too long
\end{lstlisting}

\end{document}